U-shaped Development of Newtonian Concepts:

Implications for Pedagogical Design and Research Practice

Paul J. Camp

Spelman College

Author Note

Paul J. Camp, Department of Physics, Spelman College

This research was supported in part by a grant from the National Science Foundation

Correspondence concerning this article should be addressed to Paul Camp, Department of

Physics, Spelman College, 350 Spelman Lane, Atlanta, GA 30314

pcamp@spelman.edu



Abstract

In the years 2002-2004, the Learning by Design™ group at Georgia Tech conducted an investigation of the development of qualitative Newtonian concepts as a function of time. I describe the classroom context and method for this investigation, involving the coordination of multiple measures including diagnostic quizzes, structured interviews, ethnographic observations in class and performance assessments. This paper specifically concerns the results from the diagnostic quizzes, relying on other measures as supporting evidence. I present convincing evidence of conceptual gains from that data. I also show that when looked at in more detail, there is a clear non-monotonic conceptual development process in which there is a period of apparent regression of understanding following apparent satisfactory understanding. I consider prior work on this effect from developmental psychology, and propose a hypothesis to explain its presence in this data. I then consider how cognitive models converge on a similar explanation, and in the process contrast this interpretation with the currently extant interpretation of related work in terms of an interference effect. I close with consideration of implications of this effect for both pedagogical design and research practice.



## Introduction

This report describes what appears to be the first observation of the phenomenon of U-shaped development in physics education. U-shaped development (Siegler 2004, to be described more fully below) is thought to be a general feature of learning in all but the simplest skills.

"U-shaped development" refers to the peculiar phenomenon in which learners can *appear* to forget, temporarily, what they have already learned. This goes contrary to the conventional wisdom that conceptual understanding and performance ability are tightly coupled and increase together. It appears to have escaped attention previously, perhaps because many investigations in physics education[1] have been interested more in assessing large scale intervention strategies or in small scale investigations of individual conceptual changes. The assessment strategies deployed for these purposes often do not easily afford observing U-shaped development since they were not designed for that purpose. An assessment strategy of a large scale intervention could, for example, conduct observations fairly widely spaced in time, which then might end up missing the rapid changes characteristic of U-shaped development. Alternatively, an assessment of, say, a tutorial on a narrowly defined concept may not track the student past the end of the tutorial and would then miss any subsequent performance variations. U-shaped development usually reveals itself only in situations where closely spaced observations are being conducted over an extended period of instructional time centered on the same set of concepts. Otherwise, it easily escapes detection. I was fortunate to be in a situation where the classroom context and the research tools available made the developmental progression easily visible.

---

[1] Though not, it should be noted, in some that have occurred since, notably Sayre and Heckler 2009 and Clark, et. al. (2010)

I will first describe the classroom context and curriculum structure in which the observations took place, and the assessment strategy we deployed. The pedagogical approach used was Learning by Design (LBD™), a project- and design-based, iteratively structured middle school science curriculum, has been described elsewhere in the literature. Nevertheless, I include here a brief overview because it will prove critical for interpreting the observational results for the reader to understand how LBD makes multiple spirals through the *same conceptual material* as well as to see how the assessment strategy was integrated into the instructional environment. The context shifts from one spiral to the next but the content does not. After this summary, I will describe the results of those observations in some detail, both in terms of the overall learning outcomes from the class as well as the ways in which U-shaped development is revealed by those observations. Following that discussion, I will summarize prior research on U-shaped development from developmental psychology, together with the learning processes that appear to cause it to occur and propose a mechanism to account for the data. I will approach this from the point of view of two different cognitive models: case-based reasoning and learning resources – and show that they converge on the same conclusions, consistent with the developmental interpretations. During this discussion, I will consider in some detail the interference explanation that has been suggested elsewhere to account for similar data. I will then look at structured interview transcripts taken from the same time as the quiz data for evidence of the cognitive processes I will suggest. My hypothesis suggests that U-shaped development should be expected to be a general feature of learning and not an isolated phenomenon. In particular, I will endeavor



to show that it is more general than would be suggested by the interference explanation. Finally, I will draw some major and direct implications of the existence of U-shaped development for both pedagogical design as well as research practice.

One further note: there are two sections of this paper that are informed opinion based on the data presented. These sections are clearly marked as "Reflection" and should be taken as intended to generate further discussion, not as statements of fact.

## Classroom Context for Observations: Learning by Design™

Beginning around 1996, the Learning by Design (LBD™) team at the Georgia Institute of Technology began a concerted effort to use cognitive models of learning and memory as design guidelines for science curricula. Specifically, Case-based Reasoning (Kolodner 1993, Schank 1999), a model of learning and reasoning from analogy with prior experience which has an explicit model of memory storage for transferable learning (Kolodner et. al. 2003), was merged with Problem/Project-based learning (Barrows 1985, Koschmann et. al.1994), a cognitive apprenticeship approach to learning, to create project-based middle school science curricula in physics and geology.

LBD (see Kolodner et. al. 2003 for a description of the process by which LBD was developed) was based on the use of engineering projects as a context in which to learn scientific concepts. But in addition to the specific curricula that resulted, general design frameworks and classroom orchestration procedures were created that have much broader applicability. LBD has been evaluated with some two dozen teachers and 3500 students. It has been shown to result in significant gains in collaboration and general scientific reasoning skills (Kolodner et. al. 2003). This paper will, in part, report equally significant gains in



physics content learning. In short, LBD promotes not only conceptual knowledge but also deep learning of scientific practices and explicit reflection on productive learning strategies at the same time.

The overall design of any LBD unit involves placing students right in the middle of a complex challenge from the very beginning, together with all the relevant scientific concepts needed to work through the challenge. Instead of developing their understanding in a linear fashion, progressing in a logical deductive sequence from, say, basic kinematics to vector kinematics to Newton's Laws to application of Newton's Laws, they are wrestling with all these ideas simultaneously and making multiple passes through them in shifting contexts to iteratively refine and generalize their understanding. Rather than achieving a complete understanding of acceleration before proceeding to force, LBD affords the use of an evolving understanding of acceleration to assist in the development of an understanding of force, and vice versa. Students work from the center outwards, rather than being set on an artificial and predetermined path.

In addition, LBD leverages distributed expertise (Brown et. al. 1993) as a means of developing communication, collaboration, critical thinking, and science practice skills. Every project divides, during class discussion, into a set of sub-investigations. This means that each group is working on a different piece of the overall challenge, but every group needs to understand the work of every other group in order to succeed in the challenge. A significant amount of class time is given over to practices that support learning from the experiences of others. See Holbrooke et. al. 2000, and Kolodner et. al. 2003 for a more complete description of these processes and their assessment.



LBD uses the practices and artifacts of Problem-based Learning (such as whiteboards[2]) as organizational tools, though it expands them radically. There are whole class whiteboards as well as individual whiteboards. There are Design Diaries to reflect on lessons learned, plan future activities, and assist in decoding and applying things such as expert case histories. Students use Rules of Thumb as intermediate abstractions, positioned between direct experiential results and abstract principles of science, and tools exist to help them make connections in both directions. All of these artifacts make not only the knowledge explicit but also the process by which that knowledge was acquired. Ideas and research questions are raised, observations summarized, plans revealed, and all of them modified, discarded, enhanced over time – but always accompanied by a reason for that decision that fits into an overall plan and always obtained by a means that is modeled on actual practices of scientists and designers. At every stage, the work is driven forward not by the curriculum or the instructor but by issues and ideas raised *by the students*. The trick, then, is to put them in situations that afford raising productive questions and having some idea of what those questions are likely to be so that the instructor can be prepared for them.

All of this activity is embedded in a nested set of learning cycles summarized in Figure 1. There are two subcycles, one focused on scientific investigation and the other on application to real phenomena. Issues resulting from design outcomes (especially surprises or failures) motivate questions for experimental investigation, and fundamental results from experimental investigation in turn motivate application to the evolving designs. This learning cycle is posted in the classroom so that students always have a map in their minds of where they currently are and where they can go next. At various points in the learning cycle, a recurrent set of classroom activity structures occur in which students share results and discuss one another's ideas. Borrowing of ideas is *strongly* encouraged and supported in an LBD classroom, as long as proper credit is given, as is critical analysis of the work of others, as long as it is in the spirit of collaboration, improvement and moving ahead. For example, in the design cycle, students present their design plans for critiquing and improvement, prior to building anything, in a Pin-up Session. After testing their designs, they engage in another presentation activity, a Gallery Walk, where they demonstrate what they have built, describe the results, and answer questions. It is in these repeated, reflective and collaborative activities that scientific process skills and professional behavior practices are developed and situated in appropriate contexts.

---

[2] In problem-based learning, whiteboards (also sometimes known as KWL boards) are public record keeping devices which serve two purposes: they record the evolution of understanding for the problem, and they provide a means of generating new questions which feed into planning subsequent investigations.

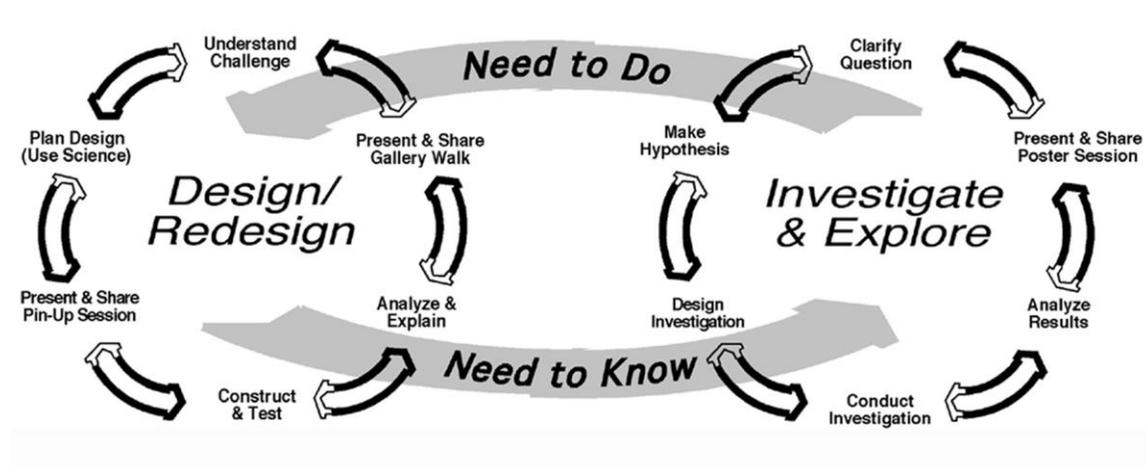

*Figure 1:* The LBD Learning Cycle

In this paper, I will concentrate on Newtonian conceptual reasoning in the physics units. There are two physics units in LBD. The first is what is called a Launcher Unit which is focused on the development, in context, of the elements of controlled variable experimental



design, assessment of uncertainty, the LBD learning cycle, and the activity structures embedded therein. Some science content is learned from the launcher activities, but it is not the primary focus and there is no systematic mention of Newtonian thinking. Coming out of the launcher unit, students understand how to behave as scientists behave, and how to create and carry out a controlled variable experiment. It is more about building a culture than about content. This occupies approximately four weeks.

**Activity Sequence for LBD: When Are Major Concepts Introduced?**

The main physics unit is called *Vehicles in Motion*. It occupies about 8-10 weeks of class time. The design challenge is to build powered vehicles that are capable of traversing a specified test track. This is the context in which Newtonian reasoning is learned. The design of the unit is fundamentally iterative. In other, words, Arons' (1990) notion of "spiraling back" is *the central design feature of the curriculum*, not just something that happens opportunistically as the occasion presents itself. The unit consists of the following parts:

1. *Messing About* In a learning cycle that has no obvious entry points, getting started is a challenge. In a curriculum that is driven forward by students' questions, helping them to raise productive questions is also an issue. LBD solves these problems through a Messing About activity in which students gain informal experience with the phenomenon. The challenge is introduced at the start, and students are encouraged to bring in toy cars from home and spend some time comparing and contrasting their construction and behavior. Following this is the first whiteboarding session, in which they describe what they've learned, speculate on how it might apply to the



challenge, and generate research questions. This activity does not explicitly introduce any major physical concepts. It is intended to explore the design space and generate useful questions for future exploration.

2. *Coaster Car* Early versions of LBD started with an open design space, but that was too large for students with limited experience to handle. Redesign was found to be more productive. They are given a minimally functioning design (a piece of foamboard, one inch diameter wooden spindles for wheels, threaded rod for axles, soda straws for bearings). An intermediate challenge is introduced, to make the car roll down a ramp and go far and straight. In the process, they investigate gravity as an "engine", explore free body diagrams and net force, and discover and apply first and second law relationships, and friction.

3. *Balloon Car* The first engine that students build is a simple balloon. This is where Newton's Third Law is introduced, completing the set of concepts for the unit. All the physical concepts raised in Coaster Car are also revisited here, in the context of a different kind of engine. It is important to remember, for subsequent discussion below, that *no new concepts are introduced after this point*.

4. *Rubber Band Car* A second engine design is introduced, a rubber band with one end wrapped around the rear axle and the other fixed to the front of the car. A second trip through all of Newton's Laws occurs at this point, with the context again shifted. The engine is such high torque that, initially, the wheels spin out and it goes nowhere. It obviously has a lot of power but does not convert that power efficiently into motion. This motivates returning to an investigation of friction and the role it plays in rolling motion. It also motivates investigation of the equivalent of gearing



systems. Students develop the idea that friction can sometimes be a good thing and begin gluing bits of rubber band on the outside of their tires, or the idea that the size of the wheel compared to the size of the axle is an important issue and giant wheels begin to appear. At every stage, they must use Newtonian reasoning in ever longer chains to account for what they are seeing.

5. *Hybrid Car* In early versions of LBD, a falling weight engine was investigated in the same level of detail as the rubber band. This activity was dropped in the interest of time, being retained only as a demonstration and class discussion leading into the hybrid design. The goal here is to make the best possible design to achieve the challenge. They usually start with a "more engines is better" idea, but through experimentation and Newtonian analysis realize that this is not the case. They usually settle on a single engine, though they may retain structural elements from other designs (for example, passing a very long rubber band over the top of the tower from the falling weight design).

Note a feature of LBD that will be critical for future discussion: all Newtonian concepts are introduced no later than the Balloon Car activity. There are no additional concepts past this point. The unit is structured to make multiple passes through the same set of concepts in variable contexts, rather than continually adding new concepts in a linear sequence.



**Assessment Strategy for Tracking the Development of Newtonian Conceptual**

**Understanding**

In the 2002-2003 and 2003-2004 academic years, we deployed a multiple measures observational approach in several LBD classrooms. The iterative nature of the classroom activities, and the fact that they were making multiple passes through Newton's Laws, afforded periodic measurements of their understanding on a fixed set of concepts. These were made pre, post, and at the end of each cycle of activities. Thus, we were able to track the development of their understanding over time. We performed the following measures:

1. *Multiple choice conceptual evaluations* These were created using age-appropriate items drawn from the Force and Motion Conceptual Evaluation (FMCE; Thornton 1995) and Force Concept Inventory (Hestenes et. al. 1992) as well as custom items created using the same developmental stages that underlie the FMCE. The questions were rated for internal purposes as easy, medium and hard through negotiation between an expert middle school teacher (drawing on experience with what students find difficult during instruction) and a physicist education researcher (drawing on descriptions of developmental processes in the literature). The full one hour evaluation was given pre and post. Due to time constraints imposed by the schools, smaller subsets of the quiz items were given after Coaster Car, Balloon Car and Rubber Band Car, lasting about 20 minutes each, and containing a mix of the various concepts and difficulty levels. Our goal in the quizzes was to measure each concept as frequently as possible while maintaining the same average difficulty level and



avoiding test/retest effects. This data is the primary focus of this paper. The full set of questions can be made available upon request. I do not make it available here in order to preserve the integrity of the FMCE and FCI.

2. *Structured Interviews* We chose a smaller subset of target students, about a half dozen from each of two teachers, for intensive interviewing. The individual interviews lasted up to an hour and students were pulled out from classes at the end of each activity cycle to participate in them. Each interview was based on a simple experiment similar to but not identical to the ones they were doing in class. The interview structure was based on gradually more specific probes to determine the level of specificity needed to prompt a Newtonian explanation. The first question was always vague and open ended, to see what they reveal. Usually, it was along the lines of "explain why that thing did what it did." If a Newtonian explanation was forthcoming, we moved on to the next question section. If the answer was non-Newtonian, then a series of more and more explicit prompts would attempt to elicit a Newtonian explanation, if the student was capable of providing it. For example, we might follow up with "how would Newton's Laws explain this?" and if that didn't work, "How would you use Newton's Second Law to explain it?" and if that failed, we might start inquiring about individual forces acting on the particle.

3. *In class ethnographic observations* Approximately three times each week, we would directly observe and video record several entire class periods to gauge what the students and the teacher were engaged in and the nature and content of the discussions that were going on. These records document the same class section across time and were the same classes from which interview participants were pulled.



4.  *Performance Assessments* Pre launcher unit, post launcher unit and post main unit, we would take over several classes for one day to stage a mini-challenge. Students had to design and carry out an experiment to solve a challenge, and justify their designs and interpret their results using scientific reasoning. These were small group assessments and each group was video recorded. The previously mentioned results on collaboration and general scientific reasoning skills largely come from coding this data (Kolodner et. al. 2003).

I will analyze in detail the results of the multiple choice conceptual exams from the 2002-03 academic year. In the interest of space, detailed analysis of the 2003-04 year will be omitted. The results are not materially different so going through all of the details would not add value. I will propose an explanation for the observations and then turn to the other lines of evidence to see the extent to which they support that interpretation.

## Analysis of Pre/Post Conceptual Evaluations – LBD vs. Comparison

The conceptual evaluations are analyzed at two levels. One level is a gestalt measurement. I select only the pretest and posttest and look at gains on the entire evaluation in an LBD class compared to a carefully matched comparison class taught in a more traditional format. Then, using data from only the LBD classroom, I pull apart the questions into distinct concept clusters and examine their development across time[3].

---

[3] Since I intend to criticize pre/post measures later in the paper, it is legitimate to object to my use of them at this point. I will answer this objection later in the paper by correlating with other measures.



The target and comparison classes are described in much more detail elsewhere (Kolodner et. al. 2003). Briefly, we compared an LBD honors science class to a non-LBD honors science class in the same school. Students were assigned to the classes by the school administration, not by us. The two teachers had comparable backgrounds and experience levels and, as will be seen by the pretest data, the student populations were comparable as well. The non-LBD classroom was not entirely lecture based, but activities tended to be much smaller scale, more directive in approach and of significantly shorter duration (usually a matter of a few days at most). The results are summarized in Table 1 and Figure 2.

Table 1

*Pre and Post Results for LBD and Comparison Classes*

|  | *2002-2003* | | | *2003-2004* | | |
|---|---|---|---|---|---|---|
|  | *Pretest (%)* | *Posttest (%)* | *Normalized Gain (%)* | *Pretest (%)* | *Posttest (%)* | *Normalized Gain (%)* |
| *LBD N=90 (02-3), 82 (03-4)* | 37 ± 7.3 | 76 ± 8.4 | 61 ± 12 | 34 ± 6.6 | 68 ± 12 | 52 ± 21 |
| *Comparison N=84 (02-3), 85 (03-4)* | 30 ± 8.5 | 39 ± 9.7 | 11 ± 20 | 28 ± 8.7 | 36 ± 7.7 | 11 ± 16 |

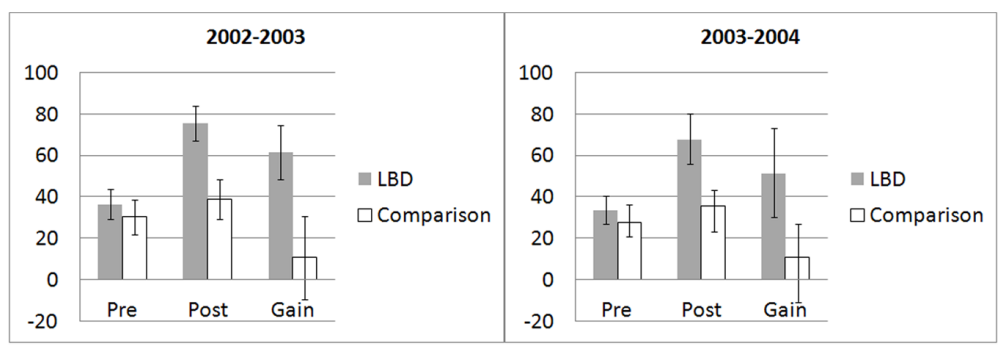

Figure 2: *Pre and Post Comaparison between LBD and Comparison Classes*



A T-Test was performed pre, post and on the gains. We draw from this data three lessons:

1. The initial performances of the two groups of students are equivalent within the standard deviations of each and with p < .001. The post tests are not equivalent, again with p < .001.

2. The LBD classes experienced a large gain, whereas the comparison classes experienced a minimal gain, consistent with zero. Again, p < .001. The effect size is quite large: 0.83 by Pearson's *r* and 3.0 by Cohen's *d*.

3. The post test scores are comparable to the best of other, non-project based interactive engagement methods but in addition, as reported in previous papers, LBD students *also* experience significant gains on non content related science practice, metacognitive awareness, and collaborative skills such as self-checks, experimental design, distributed efforts, negotiations of common understanding and use of prior knowledge (Kolodner et. al. 2003).

**Developmental Sequence of Individual Newtonian Concepts in an Iteratively Structured Classroom**

The iterative nature of an LBD classroom affords a more refined look at the development of Newtonian performance than has often been the case in physics education research. In particular, we were able to make measurements of intermediate states of performance at the end of each cycle of the *Vehicles* unit and examine the evolution of their performance over time. (Note: I use the word "performance" rather than "understanding"



quite deliberately, for reasons that will become clear below.) This data comes exclusively from the LBD classroom and does not include comparison students.

Items from the quizzes were grouped into specific concept clusters:

1. *First Law* questions that had to do with situations when $F_{net} = 0$.

2. *Second Law* anything involving nonzero net force.

3. *Third Law* questions regarding interactions between particles

4. *Net Force* questions specifically to do with how forces combine

5. *Acceleration* questions that distinguish acceleration from velocity, for the most part.

As a first cut at analysis, I look at the percentage of correct answers for each concept cluster as a function of time. First, consider results for questions involving Newton's First Law:

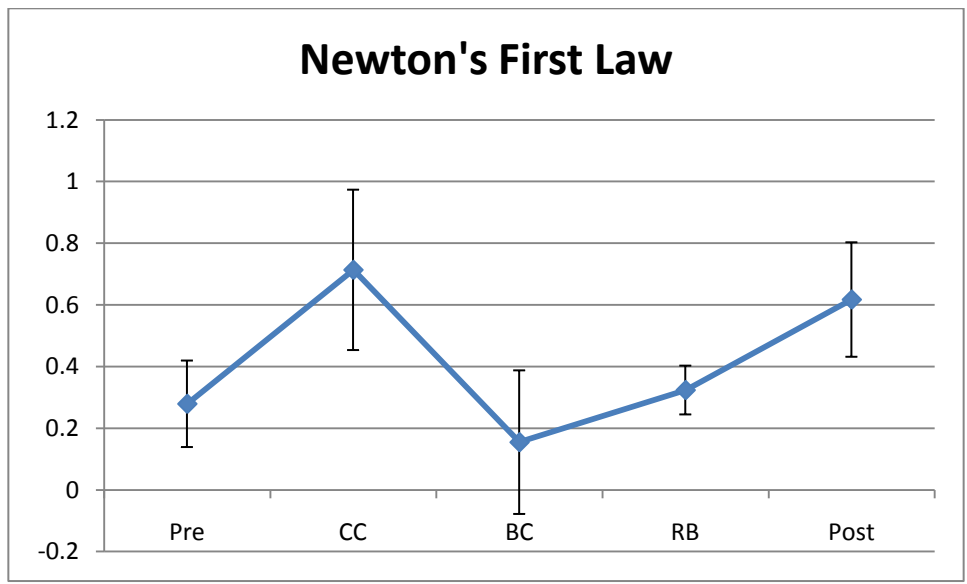

*Figure 3.* Newton's first law performance as a function of time



A repeated measures ANOVA was performed on this data revealing a significant dependence of performance on test time with p < .001, $F(2.852, 199.665) = 111.166$. Tests for significance were also performed for sequential paired samples of measurements with the result that significant differences existed with p < .001 for each of Pre to CC, CC to BC, BC to RB and RB to Post.

Next, consider data from the Second Law cluster:

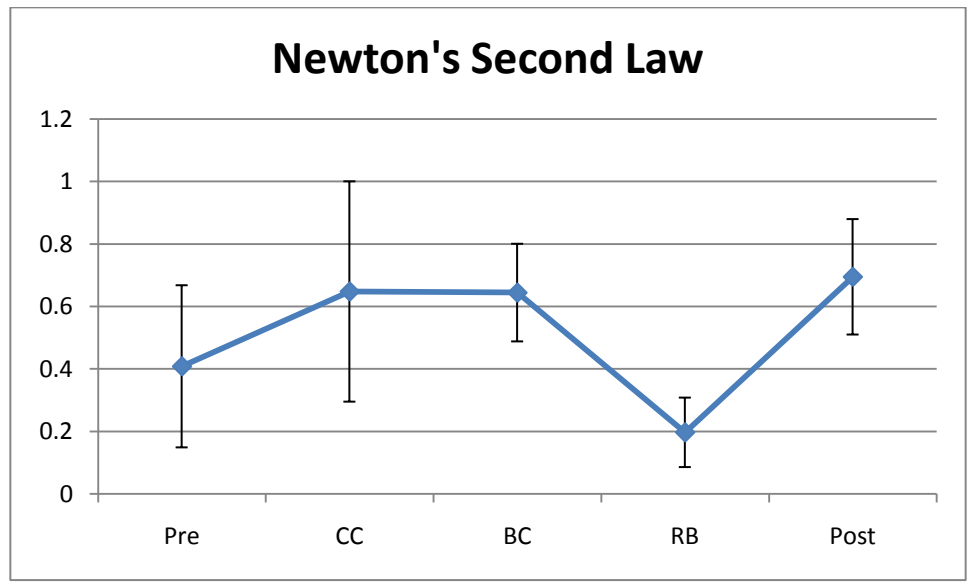

*Figure 4*. Newton's second law performance as a function of time

Again, a repeated measures ANOVA revealed a significant dependence of performance on test time with p < .001, $F(2.645, 185.130) = 66.833$. Paired sample tests revealed significant differences with p<.001 for Pre to CC, BC to RB and RB to post. There was no significant difference from CC to BC.

For the Third Law concept cluster:



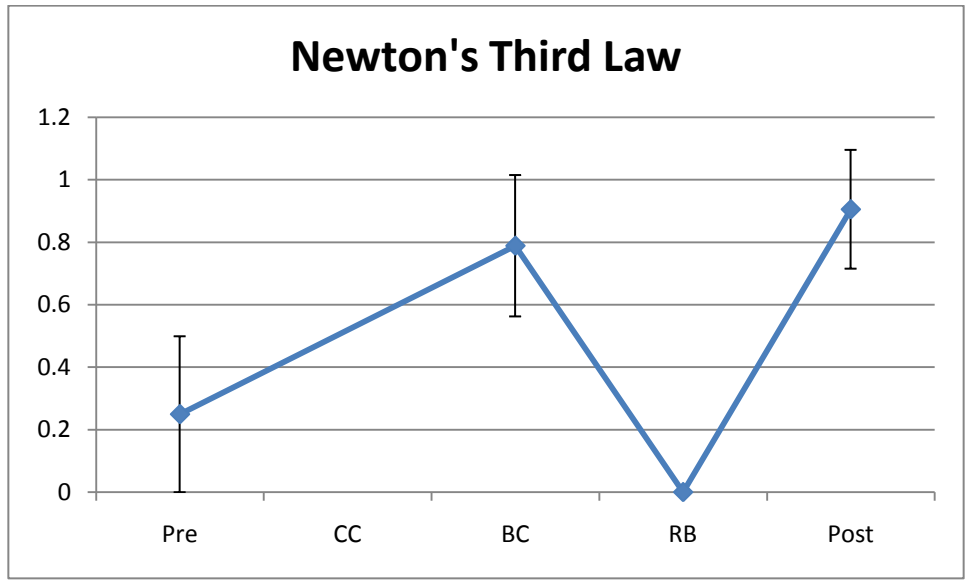

*Figure 5.* Newton's third law performance as a function of time.

Repeated measures ANOVA revealed a significant dependence of performance on test time with $p < .001$, $F_{(2.371, 165.953)} = 649.096$. Paired samples tests showed significant differences with $p < .001$ for pre to BC, BC to RB, and RB to post. Note the drop to zero at RB. This is a true zero with no uncertainty since *every* student got *every* question wrong. There was no test of third law at CC since the concept was not introduced until BC.

Two auxiliary concept clusters were also evaluated: acceleration and net force. First acceleration:



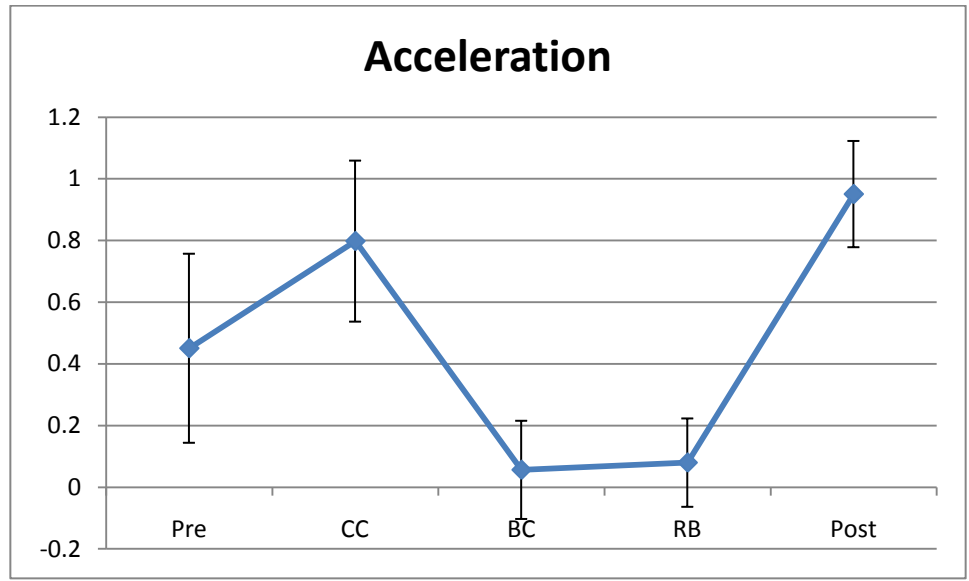

*Figure 6.* Acceleration concept performance as a function of time.

Repeated measures ANOVA revealed a significant dependence of performance on test time with p<.001, $F(3.282, 229.744)=253.896$. Paired samples tests revealed significant differences with p<.001 for pre to CC, CC to BC and RB to post. There was no significant difference for BC to RB.

And finally, the net force data:



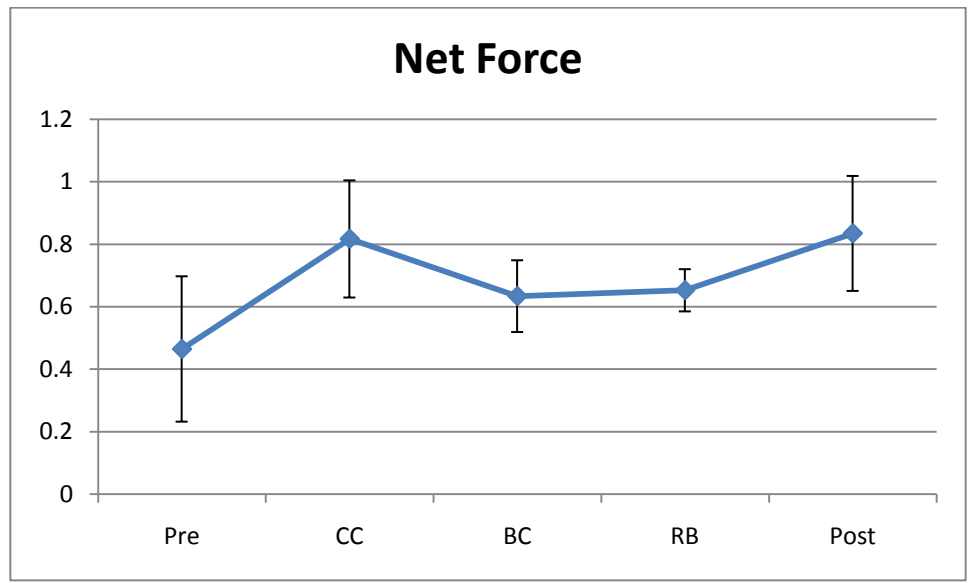

*Figure 7.* Net force concept performance as a function of time

Repeated measures ANOVA showed a significant dependence of performance on test time with $p<.001$, $F(2.703, 189.223)=60.584$. Paired samples tests revealed significant differences with $p<.001$ for pre to CC, CC to BC, and RB to post. There was no significant difference from BC to RB.

The most notable pattern by far that emerges from this data are the significant performance declines along all conceptual dimensions at intermediate points. First Law, acceleration and net force decline at BC (quiz 3) whereas second and third law performance declines later, at RB (quiz 4). This is worth exploring in some additional detail, and accounting for it will be the main focus of this paper. I will approach an explanation along multiple dimensions, and will develop an account that is at least consistent with the quiz data, as well as with concurrent structured interviews and in-class ethnographic observations. The data from 2003-04 is not materially different in any essential way from this data so I will not include the details of its analysis.



**Assessment of Question Difficulty as a Contributing Factor**

Before proceeding to that, however, it is worth addressing whether or not the performance declines were due simply to question difficulty. For this purpose, let me tabulate the questions and their difficulty level across time. Questions were balanced as much as possible to have similar difficulty levels on each concept, given the 20 minute quiz time frame. In the tables that follow, H, M and L label the difficulty level (high, medium and low) and the number that follows designates a specific question. Numbers in parentheses are the number of correct responses to that item on that quiz. I have also indicated where performance declines and increases occurred.

Table 2

*Question Difficulty Tracking Across Quizzes -- Acceleration*

|  | *Pre* | *CC* | *BC* | *RB* | *Post* |
|---|---|---|---|---|---|
| *Acceleration* | A-L1(16) A-M1(48) | A-L1(65) A-M1(51) A-H1(54) | A-M1(0) A-H1(8) *Decline* | A-L1(2) A-M1(2) A-H2(13) | A-L1(68) A-M1(67) *Increase* |

For the acceleration cluster, clearly performance variations cannot be attributed solely to variations in item difficulty across quizzes. To be sure, the decline occurs when a low difficulty question is dropped, but correct responses also dropped dramatically on the two questions that did not change between CC and BC. Since it may be argued that these questions' difficulty levels could have been misidentified, they are included in an appendix.



Table 3

*Question Difficulty Tracking Across Quizzes – Newton's First Law*

|  | **Pre** | **CC** | **BC** | **RB** | **Post** |
|---|---|---|---|---|---|
| ***1ˢᵗ Law*** | N1-L1(5)<br>N1-L2(62)<br>N1-M1(18)<br>N1-M2(6)<br>N1-H1(9)<br>N1-H2(19) | N1-L2(68)<br>N1-M2(41)<br>N1-H2(43) | N1-L2(0)<br>N1-M2(22)<br>*Decline* | N1-L2(68)<br>N1-M2(1)<br>N1-H2(0)<br>*Increase* | N1-L1(30)<br>N1-L2(68)<br>N1-M1(32)<br>N1-M2(48)<br>N1-H1(16)<br>N1-H2(69)<br>*Further Increase* |

As with acceleration, first law performance declines when a high difficulty question (N1-H2) is dropped and recovers when it reappears, but at the same time there are dramatic drops and then recoveries on low (N1-L2) and medium (N1-M2) difficulty items that were the same from CC through post. Furthermore, a high difficulty item with respectable responses at CC has no correct responses at RB and recovers to respectable levels at post. Again, one cannot consistently explain this result based on question difficulty since the *same* low, medium and high difficulty questions accompanied both the performance decline and the performance recovery. Question N1-H2 has also been included in the appendix.

Table 4

*Question Difficulty Tracking Across Quizzes– Newton's Second Law*

|  | **Pre** | **CC** | **BC** | **RB** | **Post** |
|---|---|---|---|---|---|
| ***2ⁿᵈ Law*** | N2-L2(54)<br>N2-M2(17)<br>N2-M3(16) | N2-L1(44)<br>N2-M3(48) | N2-L2(69)<br>N2-L3(69)<br>N2-M3(6)<br>N2-H1(39) | N2-L2(2)<br>N2-M1(1)<br>N2-M3(2)<br>N2-H1(51)<br>*Decline* | N2-L2(68)<br>N2-M2(15)<br>N2-M3(65)<br>*Increase* |



With Newton's Second Law, the performance decline occurs when a medium difficulty question (N2-M1) replaces a low difficulty (N2-L3) one. However, performance also declines dramatically on the low difficulty question that remains the same between the two quizzes (N2-L2). Again, change in average question difficulty is unable to account for the entire variation. It is quite interesting that, on the RB quiz, performance is notably higher on the high difficulty question than on the low and medium difficulty ones. I don't think this means that the questions have been miscategorized since performance on the BC and CC quizzes support the categorization. Also, the N2-L2 question involves comparing situations when there is only a mass change, and N2-M1 concerns a situation when the force is directed opposite to the velocity. Question N2-H1 involves considering both of these effects simultaneously. This may be evidence of a test-retest effect on a question that stood out to the students.

Table 5

*Question Difficulty Tracking Across Quizzes – Newton's Third Law*

| | Pre | CC | BC | RB | Post |
|---|---|---|---|---|---|
| *3ᵈ Law* | N3-L1(7)<br>N3-L2(69)<br>N3-L3(4)<br>N3-M1(11)<br>N3-M2(31)<br>N3-M3(1)<br>N3-H1(1) | *No Third Law Questions* | N3-L4(71)<br>N3-L5(49)<br>N3-M4(55)<br>N3-H2(49) | N3-L4(0)<br>N3-M4(0)<br>N3-H2(0)<br>*Decline* | N3-L1(65)<br>N3-L2(71)<br>N3-L3(56)<br>N3-M1(66)<br>N3-M2(66)<br>N3-M3(62)<br>N3-H1(64)<br>*Increase* |

For Newton's Third law, there are major performance declines on the same three questions (N3-L4, N3-M4, and N3-H2) between BC and RB. What changes is dropping a low difficulty question (N3-L5). Students did perform well on this question on the BC quiz,



but they performed at least as well or better on the other questions. Average question difficulty cannot account for these performance variations either. Questions N3-L2, N3-M1 and N3-H2 are included in the appendix.

Table 6

*Question Difficulty Tracking Across Quizzes – Net Force*

|  | *Pre* | *CC* | *BC* | *RB* | *Post* |
|---|---|---|---|---|---|
| *Net Force* | NF-L1(45)<br>NF-M1(26)<br>NF-M2(8)<br>NF-H1(53) | NF-L2(59)<br>NF-L3(45)<br>NF-L4(69)<br>NF-M2(0)<br>NF-M3(69)<br>NF-H1(48) | NF-M2(66)<br>NF-M3(69)<br>NF-H1(0)<br>*Decline* | NF-L2(0)<br>NF-M2(68)<br>NF-H1(71) | NF-L1(52)<br>NF-M1(47)<br>NF-M2(69)<br>NF-H1(69)<br>*Increase* |

In the Net Force cluster, essentially the entire (small) performance decline at BC is attributable to a simultaneous decline on a high difficulty item (NF-H1) and improvement on a medium difficulty item (NF-M2) which nearly compensate. In addition, several low difficulty items were dropped. Here, the performance decline may be attributable to the dropping of the low difficulty items. The best that can be said, then, is that no performance variation was detected for Net Force except for the initial increase from Pre to CC. However, there is still an intriguing hint of a U-shaped phenomenon on question NF-L2, on which students were successful on the CC quiz but everyone got wrong on the later RB quiz. Since net force is necessarily a composite concept, this may reflect multiple developmental processes occurring simultaneously. This is a point that I will return to below when looking at the patterns of incorrect answers. From the incorrect answers, we will see that the



performance decline at BC (quiz 3) is due to a single reasoning mode, so it is at least arguable that the performance decline is real and not an artifact of variations in question difficulty.

The repeated measures ANOVA suggests random variations are an unlikely explanation of this pattern. Variations in question difficulty from quiz to quiz now also seem unlikely for all concepts except Net Force. We are compelled to look elsewhere for explanations.

**Within-Test Correlations on Concept Clusters Over Time – Patterns of Incorrect Responses**

Up to this point, my discussion has been somewhat abstract. Now I will try to make it a little more concrete by describing some of the specific reasoning patterns exhibited on the quizzes and what they might indicate about how student conceptions evolve over time. All of the quizzes we administered were constructed using the developmental stages for Newtonian reasoning identified by Thornton (1995) and the pre-Newtonian reasoning patterns described by diSessa (1993). This means that the pattern of *incorrect* answers given by students is as, if not more, informative than their correct answers since it will reveal some information on intermediate states of conceptual development. Presenting the raw data would be uninformative in the absence of the full set of quiz questions so I will present summary descriptions of the analysis. Crosstabs were constructed for all answers for each concept cluster on each quiz. From the crosstabs for each quiz, every unique response pattern for all the concept clusters was extracted. This allowed us to determine how many students gave each individual response pattern, and to extract from each pattern, where



possible, what aspect of that particular concept seemed to be causing difficulty based on the incorrect responses provided by students.

**Third law.** Since the most significant effect is clearly the third law performance decline, I will address it first. The pretest showed twelve distinct patterns of answers with no significant preference for any particular one. However, some individual questions revealed incorrect responses that were common across all the patterns. Most students appeared to reason in a way consistent with the assumption that only active objects (that is, those with an internal motive force such as cars or people) can exert forces. This would, in Thornton's sequence, qualify as a student view 0 since it is a lower level of performance than all of the views he identifies. There was little variation on the answers to questions involving this mode of reasoning. Also, almost all seemed to reason that "same velocity" implies "same force" regardless of mass. Only two students did not answer consistent with this understanding.

At BC (quiz 3), most students answered mostly correctly. This quiz followed the curriculum cycle in which the third law was introduced. The only significant difficulties occurred on questions involving rockets, where 11 students responded consistent with the believe that they push off of the air and 5 with the belief that they push off of the Earth, but both groups are otherwise correct on all other questions. Also, 5 students appeared to still be aligning force with speed but, interestingly, agreed with a statement of the third law in abstract terms.

The performance decline at RB (quiz 4) is quite interesting. This quiz came at the end of a curriculum cycle in which students were applying all of Newton's laws in the context of a rubber band powered engine. They had to consider third law interactions



between the rubber band and the axle as well as between the wheels and the surface. Though every student answers every question incorrectly, they are not incorrect in the same way as they were on the pretest. In this quiz, 85 of the 90 students tested gave identical patterns of answers at this point, compared to a dozen distinct patterns with no particular preference on the pretest. As in the pretest, their reasoning was consistent with the idea that only active objects can exert forces, which had appeared to vanish at BC (quiz 3). However, all students retained their understanding of the "oppositely directed forces" condition of the third law which had first appeared on BC (quiz 3), something that was not evident on the pretest. In other words, their answers on RB (quiz 4) are consistent with an understanding in between those exhibited on the pretest and BC (quiz 3). This indicates a return to an intermediate prior state of understanding, but one that was still developmentally advanced over the pretest state.

The post-test shows a marked improvement over the pretest and over RB (quiz 4), not only in terms of overall correctness but also in that there are only two distinct patterns of answers compared to 12 on the pretest. This quiz followed a third cycle of the curriculum in which students were applying all of Newton's laws, this time in the context of a personally designed hybrid engine that might combine elements of balloon, rubber band and falling weight engines. The bulk of the disagreement on this quiz is on a single question, where nine students appear to believe that whenever two objects are motionless they are experiencing the same force (an individual force, not a net force), but were otherwise correct. This is consistent with a vestigial logic aligning force with velocity though, interestingly, only in the case of zero velocity.



**First law.** This data reveals a gradually declining tendency to reason in a way consistent with the idea that force follows velocity. This corresponds well with the patterns noted on Third Law reasoning above. Notably, the pretest data reveals a significant tendency to answer consistent with the belief (in the language of diSessa, 1993) that "change takes time" and that forces "die away." That is to say, that the effect of an interaction does not appear immediately but must "build up" in some way, nor does the effect vanish immediately when the interaction vanishes but instead must gradually decline over time, that there is a lag between the cause and the effect. Neither of these points of view is evident after the pretest. That leads to an interpretation of the performance decline at BC (quiz 3) similar to the one involving Third Law above, that there is a retreat to a prior state of understanding but that it is still in advance of the pretest state of understanding – that force follows velocity but there is no time lag. At the time of this quiz, students have completed two curriculum cycles involving the first law.

**Second law.** Pretest performance is all over the map, with 17 distinct answer patterns and no clear majority for any of them. The most notable common feature amongst these patterns is an apparent belief that force follows velocity, consistent with the first and third law data.

By CC (quiz 2) and BC (quiz 3), a majority of students show performance consistent with correctly associating force with acceleration rather than velocity. At this time, students have completed one (CC) and two (BC) curriculum cycles involving the second law. The most notable difference between the two quizzes is quite a bit less variation in the incorrect answers. The decline at RB (quiz 4) appears to stem from an issue understanding acceleration. At this point, students have completed three curriculum cycles involving the



second law. On questions in which the acceleration is not explicitly given and must be inferred, there is a marked tendency to use velocity as an indicator of acceleration. This accounts for about 2/3 of student incorrect responses. That implies that there may not be an issue with second law per se, but rather an issue with acceleration (or, more precisely, that there may not be a clear demarcation between the two concepts and that they develop in tandem). In other words, students respond as if they correctly associate force with acceleration, but are unable to correctly infer acceleration in situations where it is not provided to them explicitly. This, in turn, is consistent with the fact that the acceleration cluster showed a performance decline *before* second law, which decline in turn persisted *through* the decline in second law.

**Acceleration**. There is a marked tendency on the pretest to say that objects which are slowing down are not accelerating. This reasoning pattern does not appear thereafter. What is going on during the performance decline from BC (quiz 3, two curriculum cycles involving acceleration) through RB (quiz 4, three cycles involving acceleration) is instead the technical issue of sorting through the various indicators of acceleration. Acceleration is more of a composite idea than force. Students must identify two different measures of motion, velocity change and time interval, neither of which is directly visually perceptible, and reason based on their ratio. This is a lot of machinery to get working all at once, and most of the incorrect answer patterns show either failure to choose a correct indicator (for instance, using displacement instead of time interval) or failure to execute the ratio step.

**Net force.** There appear to be two primary conceptual issues here – failure to identify forces that are not visually salient (e.g. the force exerted by a table on a book, or the force of gravity on an object that is not falling), and a tendency to align the direction of net



force with the direction of velocity. Both of these are evident on the pretest, but subsequently the first one disappears. The small performance decline from BC (quiz 3, two curriculum cycles involving net force) through RB (quiz 4, three cycles involving net force) is due pretty much entirely to the "net force goes with velocity" line of reasoning but in a very interesting way. On questions that are closely analogous to classroom experience, students answer correctly. Failures are primarily on more distant problem situations, indicating a context bound state of reasoning which disappears by the post test. This will in fact be a very interesting point for the discussion below.

One could argue that since the most dramatic performance declines occur near the Rubber Band Car activity, perhaps there is an issue with it specifically. It is, however, not clear what that issue would be. For one thing, the pedagogical structure of Rubber Band Car is identical to that of Balloon Car which preceded it, simply with a different engine. It requires continued use of Newtonian reasoning but does not introduce any new Newtonian concepts. Students are performing experimental investigations in a long-familiar format. Rubber Band Car is a pure context shift. In any case, such an argument would still have to explain how essentially all the concepts previously learned can then simultaneously be unlearned, and quiz questions having no obvious relationship to rubber band engines (such as abstract statements of Newtonian relationships), once answered correctly, now are answered incorrectly. It would also have to account for why the acceleration, First Law and Net Force concept cluster experienced performance declines several weeks earlier, at the end of the Balloon Car investigation.

Finally, there is a clear hint in this analysis that the conceptual demarcations that physicists see between mechanical concepts are not the demarcations that are



developmentally appropriate. There are several examples above, the most notable being the way in which second law and acceleration appear to be developmentally intertwined, with the second law relationship developing faster. This is a point I shall return to in the discussion below.

## Theoretical Basis of U-shaped Development – Why Does It Happen? How General Is It?

"Practice makes perfect (Camp, 1963)."

The notion that performance always improves with time, that the more you learn, the more you know and the more you know the more you can do, is as close to an accepted law of nature as it is possible to come in education research. It is deeply ingrained in all aspects of almost every pedagogical approach. It governs the linear development of ideas in the writing of textbooks and the design of courses and activities. It also governs the evaluation of all of these activities through reliance on pre/post testing. However, as a law, it appears that it may have the same scientific status as the medieval impetus theory. The unspoken notion that you never get worse as you continue to learn has to be called into question by the phenomenon of U-shaped development. So where does it come from? As we shall see, it depends on making a careful distinction between understanding and performance. My purpose in this section is to argue that U-shaped development is a general feature of human cognition, and to give some idea of the variety of mechanisms that have been shown to account for it in different contexts.

U-shaped development was first noted in the early 1970's and rapidly became an area of vigorous research. A 1978 conference (whose proceedings are in Strauss (1982) encapsulates this early work. In that conference, there were reports documenting the



phenomenon in a wide range of disparate cognitive tasks, including language development (Bowerman 1982, Mehler 1982), artistic creativity (Gardner and Winner 1982), facial perception (Carey 1982), social cognition (Emmerich 1982), ratio reasoning (Stavey et. al. 1982), musical understanding (Bamberger 1982), intuitive thinking (Schön 1982) and scientific reasoning (Klahr 1982, Wallace 1982, Siegler and Richards 1982). In addition, other research at the time documented the phenomenon in phoneme discrimination (Werker and Tees 1983), auditory localization (Muir et. al. 1979), motor development (Thelen and Fisher 1982), development of written representations (Karmiloff-Smith 1979), and transitions between levels of expertise in various knowledge domains among adults (Feldman 1994, Feldman and Fowler 1997). Interest in U-shaped development declined until a revival around 2000, which resulted in a vigorous discussion of language acquisition, concepts and categorization, and problem solving and, as a result, the creation of a theme issue of the *Journal of Cognition and Development*. Much of what follows comes out of that issue, and especially the introductory summary by Siegler (2004). It continues to document U-shaped developmental processes in object permanence and vocabulary acquisition (Gershkoff-Stowe and Thelen 2004), gesture-referent relations (Namy et. al. 2004), memory and false memory storage and retrieval (Brainerd 2004) and category formation (Werker et. al. 2004)[4].

---

[4] An early version of this paper was criticized for this list specifically, which was described as "padding." To be clear, my goal here is to demonstrate that U-shaped development is a general feature of human learning in all but the simplest matters and part of making that case involves demonstrating the wide range of situations in which it has been observed, which situations have no obvious relationship to one another save the fact that they all happen in brains.



U-shaped development is an observation, not an explanation. I have cited this whole long list of references to give some idea of the vast and diverse range of concepts and abilities in which this developmental trajectory is observed, and to argue (as others have also done, e.g. Marcovitch and Lewkowicz 2004) that it is a general feature of human cognition and learning. I have documented its existence in the development of Newtonian reasoning in the above data, but not explained it. My goal here is to first describe the general classes of explanations that have been applied previously, and then to suggest a hypothesis for the current data based on those types of explanations. Afterwards, I will turn to structured interview transcripts as a test of that hypothesis. I will spend some considerable time on this section since, from conversations I have had with many people working in physics education research, these ideas appear to be generally unknown in the field.

There is no obvious connection linking all of the multifarious observations of U-shaped development in the list above, and that argues for it being a general feature of the underlying memory formation and recall process itself. In fact, it is so pervasive that some have been led to advocate a reorientation of the entire field of cognitive development along dynamic systems lines (Gershkoff-Stowe and Thelen 2004, Thelen and Smith 1994). Digital models of dynamic neural systems have naturally produced U-shaped developmental curves as the system attempts to maintain its current state in the face of incoming information from the environment, especially when said information exhibits strong patterns of coherent covariation (Elman et. al. 2001, Kelso 1995).

The key observation here is the existence of non-monotonic changes in performance during the learning of complex tasks. The key insight is that decrements as well as increments in such non-monotonic patterns *in performance* are byproducts of underlying



monotonic improvements *in memory and cognition*. Competencies, once learned, do not disappear but they are unusually fragile while understanding reorganizes into a more mature form, and this fragility is reflected by variability in performance. But it is this internal structural and functional reorganization that characterizes cognitive development, not the sequence of states reflected in performance. Those are the outward manifestations of underlying dynamic processes. In short, achieving a new state of organization requires passage through a state of apparent disorganization (this is related to the bootstrapping problem mentioned by diSessa (1993)). The phenomenon of U-shaped development is important in this regard because it seriously constrains the space of process-based explanations in such a way that one can identify plausible internal sources of change. It does so by demanding that any theoretical explanation account for retrograde performance in all the circumstances in which it is observed.

## Categories of Explanations for U-shaped Development Patterns

Siegler (2004) identifies three broad categories of explanations that account for all previous observations of performance degradation during development. It is worth taking a moment to see if any of these are applicable here.

**1. Newly acquired reasoning processes may be overgeneralized, and applied in situations for which the new approach is inherently flawed.** For example, Siegler (1981) has studied the development of the understanding of balance for masses arranged around a fulcrum in early childhood. Children aged 3 produce essentially random performances. Since there are three options (tilt left, tilt right or balance), they have a 33% accuracy rate. By age 6, they have learned to focus on which side of the balance has more



weight, resulting in a decline to 0% accuracy when the length of the moment arm is important. By age 9, there is a return to a chance level of accuracy, but for a different reason – they do know that both factors are important but are unable to reason through problems in which one side has more weight and the other has a longer moment arm. In this case, an improved rule twice produces a decline in performance on specific types of questions.

This type of explanation does not seem to be applicable to our data simply because it is not possible to overgeneralize Newtonian reasoning. Short of atoms and galaxies, it is always correct. Decrements in performance would therefore come from, if anything, undergeneralization.

**2. A newly acquired processing approach places greater demands on the processing system than does one that is of long experience. Therefore, the greater cognitive load may result in temporary decreases in efficiency on the older approach.** One example is the use of affective vs. linguistic cues to understand conversational meaning (Friend 2001, Morton and Trehub 2001). Infants have no understanding of spoken language at all and rely entirely on affect, even when it conflicts with the plain meaning of the words. As the ability to speak and understand language appears, there is an increasing reliance on semantic over affective cues in situations where the two conflict. The new approach crowds out the old approach, and this pattern persists into grade school. However, adolescents and adults return to reliance on affective over linguistic cues when the two are in conflict, thereby enabling the unfortunate development of sarcasm in the later years.

Again, this type of explanation does not seem to apply to the data in question here. There is, no doubt, conflict between newer and older reasoning patterns, but it is *not* the case that a productive older reasoning pattern is being crowded out. The older reasoning patterns



are primarily phenomenological and decidedly non-Newtonian (diSessa 1993, Thornton 1995) and the whole point of instruction in a physics class is to deliberately crowd them out, largely by setting situations in which they are not productive. It could be argued that a prior, phenomenological, reasoning pattern that was able to get correct answers on some questions for incorrect reasons is being crowded out by an imperfectly formed version of Newtonian-like reasoning and that this accounts for the performance dip. Indeed it would, but it would not account for the first peak. Rising from low performance to high indicates that there has been some sort of conceptual change, and that gain correlates with a time at which the central focus of class discussion is on initial understanding of Newtonian reasoning. It is also not consistent with the structured interviews (described at length below) which occurred more or less simultaneously with the quizzes and which show predominantly Newtonian reasoning at the first peak, not some third, transitional reasoning pattern that still manages to work in most cases. From correlating these three different strands of evidence, it appears that they had managed to learn Newtonian reasoning patterns quite well by the time of the first performance peak. They were not relying on non-Newtonian reasoning that managed to be correct by accident.

   **3. The resource[5] in question is actually composed of multiple subresources which develop (monotonically) at different rates.** One monotonically developing

---

[5] I use "resource" and "subresource" here in the sense of Hammer, et. al. (2005) to encompass multiple cognitive items relevant to performance, such as understanding a concept and representing it appropriately in memory, ability to retrieve it as appropriate (which may involve methods of framing current experience), ability to apply it once retrieved, epistemological beliefs about what constitutes a fundamental element of explanation, and so on.



subresource may therefore temporarily outstrip another, producing multiple peaks in performance as different subresources come on line. A classic example is the production of correct irregular past tense verb forms in English (Pinker and Prince 1988, Pinker 1991). In this case, the competition is between lexical representation of individual past tense forms versus the rule based representation "verb stem + ed." During language acquisition in childhood, experience strengthens both types of representation, but the "+ed" rule develops much more rapidly, presumably due to its simplicity and much more frequent occurrence in English. This produces constructions such as "I goed to the store" and "I eated the cookie." However, as would be expected from a competing representations model, these constructions are never universally employed. There is no age at which children consistently rely on the "+ed" representation alone. They will oscillate between, for example, "eated" and "ate," and in fact past tense formations remain quite variable for some considerable time, often into middle childhood, indicating a strongly context dependent activation rather than a "crowding out."

It is in this third category of process-based explanation that I can see some possibilities for interpreting the developmental data above. It is consistent with the situational variability noted in Newtonian reasoning in prior research (Thornton 1995, diSessa 1993) where a competing representations model would provide a natural explanation. There are two obvious broad categories of resource involved in the production of physically correct Newtonian performance – conceptual understanding, and appropriate retrieval and application of that understanding. It has been noted for some time that students can produce conflicting and inconsistent explanations for the same phenomena depending on the specific context in which they must access their understanding (see diSessa 1996 for a striking



example of this). In the case-based reasoning model of cognition (Kolodner 1993), access to understanding crucially depends on links (or indices) between that memory and contextual cues. We would say that when performance is variable, the relevant memories are poorly and/or sparsely indexed and therefore mostly inaccessible except in very specific situations. The correct pedagogical approach, predicted by this model, would therefore be to shift the same content to a series of different contexts, which require drawing on the same memories in order to be successful, and thereby deliberately triggering a series of representation failures. This, in turn, would require re-indexing those memories at a more abstract level so that they are accessible in a broader range of contexts and link to more and more remote areas of direct experience. Indeed, this is the theoretical basis on which LBD was designed in the first place. A U-shaped developmental process would therefore be a *prediction* of case-based reasoning as the memory forms first in a narrow context and then its activation conditions are progressively broadened. Further, it would also predict that a Newtonian understanding is still present, just inactive due to variant contextual cues, and so appropriate prompting may reveal it.

I therefore conjecture the following model to account for the U-shaped performance data reported above. Initially, students have a largely phenomenological, and decidedly non-Newtonian, understanding in which very different rules apply in different situations and there is no global set of rules. The first result of instruction is the development of conceptual understanding, which appears to be the easier pedagogical problem to solve. The first peak in performance appears when that particular concept is an explicit target of current instruction, is forefront in the students' minds, and therefore easily accessible when they are asked relevant questions. It is indexed directly to the current classroom context, which in



turn is the context of the quiz measuring performance. But as they move on from this initial burst of experience into other problems and different contexts, they must rely on remembering the correct Newtonian reasoning and recall is seriously complicated by the fact that their older phenomenological reasoning patterns are much, much better indexed and easily accessible than their newer Newtonian ones. They must experience failure of those memories and be reminded of what they have learned more recently. This representation or framing failure produces a decline in performance. If the class is successful, then continued experience with new contexts, coupled with representation failure and systematic reminding of the need for Newtonian explanations within those contexts, will in turn improve and broaden the indexing of Newtonian reasoning memories and lead to a second peak in performance. In short, framing resources develop more slowly than conceptual resources, and case-based reasoning provides a mechanism explaining why.

There is some empirical support for this point of view not only in the general development of case-based reasoning itself, but also in the pattern of incorrect answers that was described above. For instance, when I discussed the patterns of incorrect responses on the Net Force concept cluster over time, I noted that the only difficulties students exhibited correlated with the closeness of the match between the situation of the quiz question and the situations they had directly experienced in the classroom activities. The pattern of incorrect answers over time seemed to reveal initially context-bound understanding becoming gradually more and more generalized. This is consistent with my conjecture. Further, as components of Newtonian reasoning it is unlikely that these two sets of resources are monolithic. There seems to be indications from the ways in which students produced incorrect answers that in fact Newtonian conceptual understanding divides into subconcepts



differently from a physical point of view than from a developmental point of view, and that in an intermediate state of development, accessibility can fail on some of them but not on others. For example, we saw how the decline in third law performance was associated specifically with failure along the "equal magnitudes" dimension of that skill but that the "opposite direction" dimension was retained and always accessible. We also saw that failure on second law reasoning was associated specifically with failure on the (to a physicist) independent concept of acceleration. Other concept clusters exhibited similar phenomena.

***Independent work that may support the multiple resource development point of view.*** There is some independent evidence that may be construed as support for this point of view as well. I will describe three examples.

First, in a recent paper, Lasry et al. (2011) found that the Force Concept Inventory has the puzzling property that the overall test score has high reliability whereas the individual questions do not. They describe the state of affairs thus:

> Huffman and Heller asked: "what does the FCI actually measure?" Using classical exploratory factor analysis, they examined whether there are groups of questions in the FCI that correlate with each other, which would indicate that those items measure the same idea. Their finding that the FCI questions correlated loosely led them to conclude that the FCI does not measure a single construct. Halloun and Hestenes objected to the methodology used by Huffman and Heller and asserted that "the FCI score is a measure of one's understanding of the Newtonian concept of force." Halloun and Hestenes argued that if the students' understanding of force is



not complete, then there is no reason for questions to correlate. Using a different methodology, our results show that the FCI has a high internal consistency reliability (KR-20 > 0.8) and hence support the notion that the total FCI score measures a unique construct. However, our analysis did not determine what this unique construct is. Given that Newtonian thinkers are likely to obtain a high FCI score, Halloun and Hestenes interpret the score as a measure of an understanding of the Newtonian concept-of-force.

The hypothesis I advance, in which the Newtonian force "concept" is actually developmentally a composite entity containing a large number of interacting resources, some conceptual, some not, that all develop at their own rates along their own paths starts to make some sense of this data. One may accept that the FCI as a whole measures the Newtonian force concept, while understanding that the individual questions on the FCI measure various subsets of the resources embedded in that concept. Moreover, the individual questions, which divide the force concept the way a physicist would, do not always fall cleanly on well defined resources from the developmental perspective but instead may require the coordination of several in order to answer the question. A high score on the FCI would then occur when all or almost all sets of resources are fully developed, but since individual questions rely on crosscutting sets of resources rather than individual, independent concepts, they may not exhibit consistently high performance until nearly the same point is reached. The FMCE is a little cleaner in this regard since it is explicitly based on developmental processes.



Second, in the 1980's (see, e.g., Neves and Anderson, 1981) created a knowledge compilation model that describes a detailed mechanism to account for the observation that practice on a specific reasoning task causes performance to speed up and become automatized. The basic story is that search in a fixed problem space is sped up by parallelization. This parallelization is then modeled by a compilation process, in which specific information relevant to the task is incorporated into the decision procedures used to perform the task. Initially, those procedures are general and serial and require repeated retrieval of information from long term memory, which both conspire to make the reasoning process slow. Compilation speeds up processing by combining procedures in a way that is optimized for the task (allowing more information to fit into working memory) and incorporating in them directly the information previously retrieved from long term memory. Performance speedup then occurs through no longer having to access long term memory and through application of significantly fewer, but application-specific, rules. Anderson shows that this model correctly describes actual performance, including reproducing the power law dependence of performance on time that is observed experimentally, and especially predicting successfully the observation that, with practice, people lose conscious awareness of intermediate results in the reasoning.

It also clearly produces reasoning procedures that are very much specialized for a particular task, using representations tailored for that purpose. If the task then changes, that compiled knowledge will no longer work, partly due to an imperfect fit with the original task, but mostly because, since many procedures have been compiled into few, exit points no longer exist prior to the end of the reasoning chain. Performance will then degrade again. This will last until the knowledge is recompiled in a more abstract way that covers both



situations. Clearly, this is a version of the explanation I am suggesting for the LBD results, if one assumes that phenomenological reasoning is reasoning that was compiled long ago.

Third, in a less theoretical situation, Clark and Linn (2003) described the effect on knowledge integration of repeatedly streamlining an inquiry-based middle school curriculum in thermodynamics. They assessed 3000 students using multiple choice items, a subset of 50 students using open ended narrative response items, and a case study of a single student interviewed at multiple points during the curriculum and also beyond, to the end of high school.

The streamlining process was an attempt to determine the effect of instructional time on knowledge integration. The same topics, pedagogy and supports were provided in each of four version of the curriculum. Time savings came from condensing and increasing the efficiency of activities. In the time saved, other activities on energy, light and scientific method were added.

The authors found that streamlining dramatically reduces the integration of knowledge, with performance on some of the open ended items decreasing almost linearly from 70% to 25%. There are also some intriguing hints in the graphs of their data of nonmonotonic developmental patterns, though this is not remarked upon by the authors. However, the case study does clearly demonstrate a U-shaped pattern with fewer normative responses being given at the fourth interview than at the third and more later on, as the student engages in an ongoing struggle to reconcile instructed concepts with personal experiences.

Streamlining did not reduce content or activities. Rather, it reduced opportunities for reflection, and Clark and Linn note the critical role in creating integrated knowledge of



revisiting and reconsidering ideas multiple times over an extended period in varying contexts. They note that decreasing the length by half reduced normative responses by about half and nuanced responses by about 2/3.

This is fully consistent with my hypothesis for the data reported here. Wrestling with the issue of generalizing ideas from the specific context in which they were learned is part of integrating that instructed knowledge with experiential and other instructed prior knowledge.

**Reflection on Related Work – Is It Interference?**

Approximately seven years after this data was collected, an unrelated experiment independently discovered essentially the same phenomenon (see, e.g., Sayre and Heckler, 2009 and Clark, Sayre and Franklin, 2010). These observations were based on randomized subsamples of a large lecture class to avoid test/retest effects. The subsample is then taken as representative of the class as a whole. Performance declines were observed coincident with shifts between major sections of course content (e.g., changing from electric fields to electric potentials). The assertion is made that the performance dips are caused by the content shifts via a retroactive interference effect (Bouton, 1993; Tomlinson et. al. 2009). In this section, I will argue that interference cannot fully account for the observations reported in the present paper and so, while it may be a secondary cause, it cannot be the main effect. I will also argue that the mechanism of ongoing developmental processes I have proposed above can account for all observations to date[6]. I will spend some time on this subject

---

[6] It is also worth mentioning that the Clark and Linn (2003) work described at the end of the previous section provides some independent support for the ongoing development point of view. In their individual case study, they provided evidence that the knowledge integration process for their student in thermodynamics continued



because the notion has grown in physics education research interference is the only

explanation for U-shaped developmental patterns and that is not the case.

Interference comes in at least three distinct forms: proactive, retroactive, and output-

driven, but retroactive is clearly the type being referred to here, in which new memories

inhibit the retrieval of older ones via a "crowding out" effect. The argument is that recently

learned concepts or abilities are not yet fully automatized and require significant "processing

power" to be executed. So do concepts or abilities currently being learned, and there isn't

room for both. Consequently, retrieval of older memories is impeded. Clearly, this can be

regarded as a subset of category 2, the processing method competition category, in the

taxonomy of U-shaped causal processes described in the previous section. But U-shaped

development is a much larger phenomenon with other possible explanations beyond

interference. Because there exist other categories of explanation, simply observing a

performance dip does not uniquely point to interference as the cause. Some positive

evidence must be adduced to exclude the other possibilities.

Nevertheless, some sort of retroactive interference effect was a reasonable

hypothesis at the time of the research cited above, given that in the Sayre observations the

performance declines occurred simultaneous with the content shifts (at least, at the time

resolution of the experiment), and given that they were unaware of the data reported in this

---

on well past the end of instruction in 8th grade. The integration of his instructed and experiential knowledge

continued to improve in 10th and 12th grade interviews, which seems unlikely to be the result of an interference

effect and instead more indicative of ongoing developmental processes.



paper.[7] Nothing could be more natural than to assume that two things which coincide in time are causally related, especially if there is an off-the-shelf mechanism for explaining the effect. However, Sayre et. al. do not give direct evidence of interference but rely instead entirely on the temporal coincidence between performance and content for support. It is, at present, only a hypothesis.

As a hypothesis, it presents at least three significant issues.

- It is a major extrapolation of the idea of interference. Experiments on retroactive interference appear for the most part to be relatively short term. Most involve working memory only. Those that do not, such as paired word association tests, tend to extend over no more than a day or two (unless I have missed some). In both this paper and the Sayre articles, we are dealing with phenomena that operate on much longer time scales – weeks to months. While it is not impossible that interference could account for this data, it seems to me that some positive case must be made beyond a simple assertion that it is interference because it looks like what interference would do. This is especially true when one considers examples of U-shaped development that play out over periods of years (McNeil, 2007; Siegler, 1981; Friend, 2001; Pinker, 1991) where interference between imperfectly consolidated memories is unlikely to be a factor.

- If interference were the only available explanation of U-shaped development, then it would not be necessary to formulate a positive case for it being the explanation here.

[7] This data was presented in various forms at meetings in 2004 and 2009 but this paper is its first appearance in print



However, it is not. As I have shown in the previous section, interference is a small subset of a much larger space of explanations that have been identified, all of which lead to the same performance pattern by different mechanisms, so it is not clear, simply from looking at a U-shaped curve, that it must be a result of interference. It is easy to equate U-shaped development with interference and one must take care to avoid this pitfall. U-shaped development is an observed phenomenon; interference is a candidate mechanism to account for the phenomenon. There exist other candidate mechanisms. To equate the two is rather like equating parabolic motion with gravity. The effect is not to be confused with the cause.

- Retroactive interference is a falsifiable hypothesis since it makes a specific prediction: If the U-shaped development is caused by new concepts interfering with older but still imperfectly consolidated concepts then in a situation in which there are no new concepts there can be no interference and hence no performance dip. In LBD, the last conceptual material is presented in Balloon Car. Quizzes at that point and prior demonstrate mastery of all the major conceptual material in Newtonian reasoning. There is no new content at Rubber Band Car. All of kinematics, force concepts, and Newton's laws have been in play for several weeks by this point. What happens at Rubber Band Car is that the same concepts are applied in a new context. Nevertheless, there is still a performance dip. Since mastery has already been demonstrated and no new concepts have been introduced, interference is simply not a plausible explanation. Furthermore, if interference were a major factor, one might expect that it would be seen at Balloon Car on material introduced at Coaster Car since the new Third Law concepts could plausibly interfere with the older concepts.



However, the Second Law performance is stable at Balloon Car and does not decline until Rubber Band Car. Since it appears at the time of a context shift, it must be something having to do with the transfer of existing concepts to a new context that is the underlying factor. It also seems reasonable that a correct explanation should be able to account for both these observations as well as those of Sayre, et. al. Interference only accounts for one set of observations. If another explanation can account for both, it would be the most economical hypothesis.

It is possible that there are two distinct processes going on, and that these observations that seem to be related are in fact not. I don't think that is very likely. It is important to note that (following the example from above) electric potential is not only distinctly different content from electric fields, but it also addresses distinctly different types of questions. That, indeed, is the reason for introducing it in the first place. This means that in addition to the change of content in the Sayre observations, there is also an underlying change of the context in which that content is applied. Since there is a change of content + context in one situation but context only in the other, it seems unlikely that interference, which requires new content, will prove an adequate explanation of both. On the other hand, an explanation that depends on context would plausibly function for both.

What happens in LBD is that both second and third law reasoning are being used in Balloon Car activities. The main point is to learn the third law, but the second law must be used to clarify why objects that experience the same force can respond with very different motions. By Rubber Band Car, the pedagogical scaffolding for both laws is beginning to fade. There is increasing reliance on students recognizing for themselves the need to apply Newtonian reasoning and being prompted to do so if recognition fails. Thus, it is plausible



to see the first decline on second law performance coincident with a decline on third law performance, something that would not be expected from retroactive interference.

On the other hand, the First Law plays no significant role in the Balloon Car activities, so either interference or multiple resource development would cause one to expect a decline in performance at Balloon Car and that is exactly what happens. However, again it seems reasonable to follow a law of parsimony – a single mechanism that accounts for all observations is to be preferred over one which accounts only for some.

Since the context is what changed in both the LBD and the Sayre observations, that seems to be the most reasonable causal factor to which the U-shaped performance patterns should be attributed. The case-based reasoning model provides theoretical support for this point of view since it implies that memory indices should be heavily based on the specific context in which concepts are learned, making them difficult to access when the context changes and thereby requiring a reindexing of those memories. This, indeed, was a central design principle in LBD. It points toward the multiple resource development category in the taxonomy of U-shaped explanations, as mentioned above.

We will see in the structured interview data below some direct evidence that supports this point of view. Specifically, the story from the interviews seems to be that the greater distance there is between the specific situation that students are examining and their prior experience with the LBD projects, the more difficulty they have recognizing the existence or relevance of Newtonian relationships. This is consistent with what case-based reasoning would predict. It is also consistent with the account of U-shaped development presented above. What it is *not* consistent with is any sort of interference since there is no evidence that any new concepts are taking the place of Newtonian concepts in reasoning. Students are



consistently trying to deploy Newtonian reasoning, not always successfully, or when that fails, reverting to older phenomenological reasoning patterns.

It is, of course, possible that in the Sayre observations interference is taking place in addition to multiple skill development. However, I repeat that there is no evidence one way or the other at the present time. An alternative argument might be that the students in question are having difficulty sorting through the boundaries of applicability of electric field versus electric potential representations. The fact that the performance pattern in LBD is just as pronounced as that in Sayre's work, despite interference being excluded as a possible interpretation, does suggest to me that interference, if it is present, is not the primary causal mechanism.

## Reflection on the Case-based Reasoning and Learning Resources Cognitive Models – Appreciating the Complementarity

An early reader of this paper expressed the opinion that it should use the learning resources point of view (Hammer et. al., 2005; Hammer, 2000) rather than case-based reasoning as the cognitive model for both describing the LBD pedagogy and for interpreting the resulting data. Upon reflection, I realized that, while I disagree with the specific objection, each of these points of view contribute in different ways to understanding the U-shaped phenomenon. I will first explain why I disagreed with this assessment since it seems like an issue that might occur to other readers as well. Then, I will describe what I see as the relationship between the two, and finally I will describe what I think the resources model has to say about U-shaped development.



Both Case-based Reasoning and Learning Resources are alternative models within the constructivist theory of learning, in which learning is a process of constructing and revising mental models. The Case-based reasoning model focuses on reasoning by adapting prior experience to current situations and describes that mechanism in great detail, whereas the Learning Resource model describes a much more general space of reasoning patterns and so does not contain a single reasoning process. Indeed, it often views such processes as highly idiosyncratic, tending to focus more on the empirical correlations between active knowledge elements and the contextual cues that activate them than on the mechanisms by which those specific activations occur. As models, each of them has particular strengths and weaknesses.

Physics seeks the fundamental laws of nature, operating on the assumption that there is only one such set of laws, though there may be multiple equivalent ways of approaching them. There is but one quantum theory, though there are multiple ways to represent that theory mathematically. "Theory," in the sense that a physicist understands the term does not exist in cognitive science. Instead, there are a large number of models based on empirical data and developed to reproduce or explain different cognitive processes at various levels of abstraction and various degrees of generality. As the study of learning and reasoning has become a more and more significant part of physics research, some physicists have tended to treat cognitive models as if they were theories in the physical sense.

The resource model is in part an attempt to create a general framework that embraces aspects of precursor models as special cases. From this point of view, I think it is unfortunate that the original papers on the resource model (e.g. Hammer, et. al., 2005) do not mention case based reasoning (Kolodner, 1993) as one of those precursors, particularly



since much of what the resource model considers framing seems to overlap with what case based reasoning describes as a case. Likely, this is due to the historical fact that the resource model is an outgrowth of cognitive linguistics and phenomenological reasoning whereas case based reasoning comes from the largely independent line of research in intelligent systems and digital modeling of human reasoning.

For example, Hammer, et. al. (2005), say the following:

> In our theoretical perspective, framing generally involves the activation of numerous low-inertia cognitive resources rather that a single high-inertia cognitive unit. Therefore, the resulting cognitive and behavioral stabilities are local to the moment. However, as noted, when the same locally coherent set of resources becomes activated again and again, it can eventually become sufficiently established to act as a unit.

This description is a fairly good summary of about half of what case based reasoning considers the process of forming a case. However, case based reasoning adds some value to this account in that it describes in some considerable detail a specific mechanism for achieving coherent stability, an account of the cause of inertial variations, and the sorts of prospective and retrospective reflective actions that will enhance the probability of establishing these units or cases. Many of these actions are, of course, embodied explicitly in LBD activities.

As models, they each have strengths and weaknesses, as do all models. In aiming for breadth of applicability, learning resources gives up specificity with regard to the details of the mechanisms by which local coherence is achieved as well as the ability to predict what



specific experiences will provoke local coherence to occur. Case-based reasoning, on the other hand, is tightly focused on one specific learning process – adapting prior experience to current situations. It has a great deal to say about the specific mechanisms by which memories are accessed, and how they are modified to fit new circumstances, and how to store memories to that they will be better prepared for future access. It covers a smaller range of cognitive phenomena in much greater predictive detail. On the other hand, the resource model speaks in much greater detail about the extent to which epistemological beliefs influence the selection of resources brought to bear on a specific problem. You could embody this in case based reasoning as a specific type of index (a concept to be elaborated on in a moment), but it has no particular reason to have a higher or lower priority than any other index except insofar as the same epistemological index may be attached to a large number of cases.

In my reading, the resource model, as currently described, deals to a large degree with how and why specific resources are activated and is quite a bit more vague on the issues of how a set of resources may be prepared for future activation, or how they may be activated in situations where they *almost* fit, and how they might be modified or adapted to (as they say on the golf course) "improve the lie." The concept of framing only gets you part way there since it begs the question: "How do you decide which frame to use when none of them fit perfectly?" Lacking such mechanisms, resource-based explanations tend to be post hoc. The creators of the learning resource model recognized this difficulty and that led to an attempt to define "context" in a way that, to me, has a large degree of circularity to the logic. Hammer, et. al. (2005) propose "By 'context,' for an individual with respect to a set of resources, we mean *the circumstances for passive but reliable activation* [emphasis in the original]."



To be fair, they also describe this not as a definition of context but as an approach to a definition. However, to me, it seems to say "What does a context do? When it recurs, it causes reliable activation of the same set of resources (context triggers activation . . .). How do we identify a specific recurring context? When we observe reliable activation of the same set of resources (. . . and activation specifies context)." I am sure this is not what was meant, but it indicates the difficulty of making certain types of statements when constrained by the language of a model rather than a theory. An attempt is clearly being made to define "context" with sufficient wiggle room to fit "almost-contexts," but it ends up being operationalized as "the thing you want to identify in order to predict what will happen is identified by what happens once it has occurred." This definitional difficulty seems to be the reason why some work on the resource model (e.g. Hammer, 2005) explicitly disclaims thinking in terms of transfer.

Case-based reasoning, however, is quite detailed about the means by which memory indices are used to:

- label specific resources for reliable access

- connect those resources into larger units via shared indices

- define a context, or even an approximate context, by the extent to which these indices overlap with current perceptual information

- say something credible about the probability of activation as a function of the extent of index overlap



- hint at a neural mechanism for triggering activation after approximate fits (if indices correspond to neural structures, as it seems they must, then that which is wired together fires together, a logical consequence of Hebb's postulate (Hebb, 1949))

- cue both appropriate and inappropriate resources into activity depending on the indices attached to them (just because it is an index doesn't mean it is a good one)

In addition, CBR describes in detail the experiential mechanisms that can lead to filtering, discarding, adding or adjusting indices to achieve a better outcome. In fact, this account of indexing is largely the story of case-based reasoning. In a very real sense, it is little other than a very detailed theory of transfer. If one regards a case as a type of frame, case-based reasoning goes into very fine grained detail about how to construct one. Case-based reasoning was originally created as an effort to digitally model the very human process of reasoning by analogy and experience, and both of those are often an imperfect fit to the current context, yet we manage to make them work. It is very much concerned with the details of the process of conceptual descent with modification. The resource model is somewhat vague on this point, but it seems to me that some of the perspective of case-based reasoning could easily and profitably be incorporated into it.

Because of these features, case-based reasoning is extraordinarily useful as a design framework for pedagogies that will encourage a specific type of reasoning by analogy with prior experience whereas (for me, at least) the learning resources model is primarily useful as a language within which to construct a posteriori descriptions of what happened rather than a priori predictions of what specifically should be done to make it happen. But this is only what we should expect from models, all of which are necessarily good at some things and not good at others.



In short, it seems to me that these two models are complementary, and have a great deal to say to each other, and to those of us trying to understand how to engineer and understand what is going on in learning environments. This reflection should be considered an attempt to start that conversation. The reasoning in this section is what has led me to take a multiple model approach in this paper, wherein I have used the language of resources as a means of thinking about dimensions of the conceptual space that the students are navigating, and the language of case-based reasoning as a means of understanding the underlying developmental changes they experience.

**Reflection on Cognitive Models – What Does the Resource Model Have to Say about the Interference Hypothesis?**

To continue the theme of the previous reflection, the case-based reasoning account I have given of U-shaped development is that the successive refinement, revision and abstraction of memory indices can make them, over time, a more likely fit to current contexts and therefore more likely to be activated, whereas early on, when those memories are tied to indices tightly bound to the original learning context, activation is less likely so when the context changes performance can then decline. Furthermore, if as I have suggested, the atomic elements of Newtonian concepts from a developmental point of view differ from the physical point of view and, early on, are not well integrated by a web of indices, then we can easily see partial activation of Newtonian resources with gaps being filled by older, more richly indexed phenomenological resources. That could account for the performance regression being to a prior intermediate state rather than a total loss of access. This mechanism thus provides a plausible account for why the interference hypothesis is not



a good fit to the present data. It is then worth redescribing the situation in the language of learning resources to see if that sheds some additional light.

The resource model as used in physics education most often speaks about two specific categories of resources that are brought to bear to think about problems: conceptual resources, which afford understanding of physical phenomena, and epistemological resources, which afford understanding of cognitive phenomena[8]. The latter play a large role in framing a situation so as to activate the former. Activation of an appropriate epistemological resource can lead to understanding through consequent activation of relevant conceptual resources, whereas activation of an inappropriate epistemological resource can impede such understanding or frustrate it entirely.

As will be seen in the structured interview data below, this is not a bad description of the ways in which our target students did, or did not, bring to bear their understanding of Newton's laws to explain various situations. Some of them will, for example, fail to frame a friction example in the same way as a rubber band example. It also parallels in an obvious way my working hypothesis for understanding U-shaped development: there are two broad sets of resources that develop at different rates. There are those having to do with understanding Newton's laws in the first place (conceptual resources). And then there are those having to do with understanding when to activate Newton's laws (epistemological resources). This point of view is further supported by independent work (Redish, Saul and Steinberg 1997), which shows that even when students learn Newton's Third Law as determined by questions that measure that item in isolation, they are not prepared to integrate that concept into their future learning about multiple body problem solving. In other words, they are in a state similar to the students in the present study – they have assembled a useful set of conceptual resources together with a less useful set of epistemological resources to supply activation conditions. Furthermore, epistemological resources often have to do with personal beliefs about knowledge and the nature of the world. It makes some degree of sense that these should be more stable, and more difficult and slower to change, than conceptual resources. This is a matter that bears further investigation.

---

[8] There is some burgeoning interest in social resources, but those will not concern us here.

There is some support for this point of view in the incorrect answer analysis from the quiz data presented at the beginning of this paper. Recall that the performance declines did not represent total inability to use a concept. The pattern of answers did not go all the way back to the pretest pattern. Rather, it retreated to an intermediate state of development in which only certain pieces of the concept were being used. For example, in the Third Law performance decline, students retained the "opposite directions" piece of the concept, which had not been reliably evident on the pretest, but they lost their grip on the "equal magnitudes" piece. And in the Second Law data, students retained the "force follows acceleration" part of the concept but were unable to reliably identify accelerations that they had been able to do previously.

In the multiple resources account, this makes perfect sense. The "opposite directions" piece is somewhat more visually salient than the "equal magnitudes" piece (especially if the masses differ) and so it makes sense that the former should develop more rapidly than the latter, which involves also coordinating the second law principle that the motion observed is not a direct measure of the force applied. It isn't clear, from the



interference point of view, why both subconcepts should not be interfered with equally, *unless* one is mindful of their differing states of development and assumes that interference will pick on the weaker subconcept. But in that case, interference is playing a clearly secondary role, and the primary driver of the performance decline is the differing developmental rates of the various subconcepts. That is the best case scenario but, as I have already indicated, in the particular observations I have reported here, even that explanation won't work since there are no new concepts to interfere with the old ones.

The multiple resources account does not seem to bear much relationship to an interference account. The interference mechanism entails conceptual resources competing with other conceptual resources for processing time whereas the actual issue appears to involve frame (or case) boundaries. The data I have presented, as well as that of Redish, Saul and Steinberg previously referenced tell a story of difficulty activating a fully developed conceptual skill when the context changes. This is the story of a complex ability to activate and utilize a set of conceptual resources, said activation involving epistemological resources that are incompletely acquired and impeding activation. It is not the story of a fully realized ability, incompletely automatized, and being impeded or obstructed by a different, complete (or incomplete, for that matter) set of resources.

## Structured Interview Evidence for the Multiple Resources Account of U-shaped Development

I now turn toward independent observational support from classroom observations and structured interviews of a subset of LBD students performed during the performance decline. This will provide direct evidence of their thought processes at that time.



It is important to note that in this explanation, degradation of performance is *not* caused by degradation of understanding. Competence is not something that inheres in a task. It is something that inheres in a mental representation (Marcus 2004). (This fact will be of vital importance in the sequel below.) Quite the contrary, this explanation implies that relevant knowledge is present but not reliably accessible, and that understanding is continuing to develop (in the area of increasing accessibility) even while performance degrades. That is a prediction which we can check by looking at the interview and ethnographic observations that were collected during the same time frame. While this data has not yet been analyzed completely, some patterns have emerged. For example, classroom observations have been cataloged according to surface features of the discourse and specifically with respect to the particular topics being discussed at any moment. For example, the serious decline on third law reasoning at RB in 2002-2003 is accompanied by vigorous in-class disputation over applicability of the third law to interactions between the car wheels and the whole Earth.

More useful is the structured interview data that was taken from particular target students within the LBD class at more or less the same time as each of the quizzes. Since we are interested in the state of their reasoning patterns at RB, I will present transcript excerpts from interview protocols for four students taken at that time. These transcripts are portions of much larger interviews, and have been selected to highlight Third Law reasoning, for which there is the strongest variation in the quiz data. All of these students performed well on the BC quiz, which focuses heavily on third law relationships. The context of each of these interviews is a simple set of experiments with a toy car on a plastic track. In the first version, the car is fired out of a spring powered launcher on a horizontal track and coasts to



a stop. Then, variations are tried, one with a rough material on part of the track, and another in which the spring launcher is replaced by a hill whose height is tailored to make the car travel the same distance as with the launcher. So the context of the interview experiment is quite close to that of the in-class activities.

The interviews were designed to flexibly follow a more or less rigid hierarchy of prompts. The goal was to deliberately structure the interview so as to assess the degree of accessibility of Newtonian concepts to the student, the extent to which the current context could cue them into activity. So for each phase of the interview, we began with very general, nonspecific prompts. As we determined the understanding which the student was willing to reveal, we had a tree of subsidiary prompts which we could follow, in which increasing degrees of specificity in the prompts provided a rough measure of the accessibility of a specific concept in this particular context. You will be able to see this easily in the first transcript.

The first student, whom we will call A, did not mention Third Law relationships until they were brought up in the course of the interview. In what follows, "I" denotes statements made by the interviewer.

I: What laws do you recall?

A: I think the first one is any object in motion stays in motion unless it is acted upon by another object. Third law [long pause] I don't remember third law. It's got something to do with acceleration and velocity because I remember the second and third law quiz was finding velocity from, finding acceleration from the starting velocity and the final velocity over time.



I: This semester you guys learned about forces in pairs. What do you know about forces in pairs?

A: They're the same size.

I: What does it mean? What does forces in pairs mean?

A: That two forces acting upon each other, like going against each other I guess.

I: OK, give me an example.

A: Like I know we did a lot of forces in pairs with the balloon cars, where the air was pushing out on the balloon and the balloon was pushing back in on the air which forced the air out.

I: Can you describe those two forces?

A: Um, the forces would be the same?

I: Same what?

A: Same size, like if you were to draw a net force diagram or something, the forces would be the same. And then they would get pushed out the hole because that is the only way that the air would get out and the force would get out. I guess. Yeah, I think so.

I: So you said the balloon pushes on the air

A: Yeah

I: and the air pushes on the balloon

A: Yeah

I: and those two forces are the same size. Can you talk to me about their directions?



A: The balloon pushes back on the air so it's pushing inside the balloon, and the air pushes out on the balloon.

I: In this experiment, with the car going down the track, are there any places along here where forces in pairs are demonstrated, where you see forces in pairs?

A: I don't think so but I'm probably wrong.

I: When the car is in the launcher, do you think there's forces in pairs in the launcher?

A: [quickly] Yeah, yeah, yeah. Because the car pushes back on the launcher, and the red block [the launcher] pushes on the car.

I: Can you talk to me about those two forces in terms of their size and direction?

A: Um, well, the car has the same force on the red block as the red block has on the car. [Indicates opposite directions by pointing fingers when mentioning each force].

I: You were pointing – does the car push in this direction and the launcher in that direction?

A: Yeah

I: These are opposite directions. Are forces in pairs always in opposite directions, or sometimes, or never?

A: Yeah, because like in the balloon car you had the air going in one direction and the balloon pushing back in the other direction



I: What about when the car is coasting down the track. Are there any examples of forces in pairs?

A: Um, the only force I think, um, the only force on the car that would actually affect it is the friction going backwards and that's, and nothing's pushing forwards on it so there's no forces in pairs.

A did not recognize the law by name, but he did clearly enunciate the principle and connect it with his balloon car experience. However, he is clearly struggling with applicability conditions. He spontaneously articulated the "equal size" condition but only mentions "opposite directions" with some mild prompting. He can recognize, when prompted to look, a third law relationship in the launcher situation, which is fairly close to the balloon car and collision situations he has encountered previously, but again needed to be prompted to look at that specific interaction. He was not able to do so on his own. He is also having difficulty identifying a third law force to pair up with the friction on the car while it is travelling down the track. At this point, he declares there is a friction force on the car, but no forces in pairs, even though earlier he was emphatic about the universality of third law relationships. His final statement is quite revealing about the nature of his difficulty in that he appears to be looking for paired forces acting on the car alone, whereas with the balloons and the launcher he was looking at paired forces acting on different objects. However, this is an applicability condition for the third law – it applies whenever two objects interact, regardless of whether they are moving. He appears to be cueing on motion as an indicator of when objects have been acted on by forces. In the first two cases, two things are visibly moving (air/balloon; car/launcher) but in the last case only the car is visibly moving. If the universality of the



Third Law in all interactions were a fully developed epistemological resource, we might expect him to search for something else for the car to interact with or to exhibit some confusion. He does not. He is happy to assert universality, apply it when he can see it happening and then fail to apply it when the context changes. We will see that this is a common theme in all the excerpts.

The second student, whom we will call B, also failed to consider third law relationships until specifically prompted.

I: Can you explain to me in your own words what Newton's Third Law is?

B: I don't even remember which one it is.

I: Do you know what the three laws are?

B: The first one is . . .

I: Don't worry about the numbers so much.

B: I don't remember which one the third one is.

I: Ok but you know the three laws, is what you're saying, but you just don't know . . .

B: One of them, I think it's the third one, is . . . like in the back of my mind somewhere, I can't remember it.

I: Which laws do you remember?

B: Um, that an object keeps going in the same motion unless another force interacts with it, and that acceleration equals force times, er, there's an equation in there somewhere. Force times mass equals acceleration, no, mass times acceleration equals force.



I: Do you know something about forces in pairs?

B: Forces in pairs, yes. Forces always work in pairs, and if the object isn't moving they're equal in opposite directions, and like right at the moment of impact they're exactly the same.

I: Ok, so let's talk about, are there any cases in this sequence here where forces in pairs are working?

B: Normal force and the force of gravity are always working together, and they're equal if it's not moving up or down, and . . . [trails off]

I: You've given me a vertical one. Are there any that actually are [unintelligible]

B: The launcher, right at the moment that the launcher hits the car, before it starts moving, the force of the car is pushing on the launcher and the launcher is pushing on the car with the same force

I: Those are opposite directions too?

B: Right.

I: What about on the hill? Besides gravity and normal force and all that stuff, are there any third law interactions going on as pertains to horizontal direction or along the track?

B: I don't think so. Not when they're equal, I don't think.

I: So when it's sitting on the hill, moving down the hill, you don't see any third law interactions?

B: Right because there's nothing pushing it back up except the friction, which isn't equal because it's moving.



B is struggling with a few issues. As with A, she can recall the third law if prompted but is having issues with applicability conditions. She appears to believe that lack of motion is a condition for there to be an equal but opposite relationship, a feature of her identification of normal force and gravity as a force pair, her explanation of the interaction between the car and the launcher ("before it starts moving"), and with her analysis of friction as being unpaired. She articulates an implicit argument for universality, as did A, but fails to activate that understanding for friction. For both A and B, there is a disconnect between stating an understanding of the third law in connection to previous experience, and activating that understanding in a new experience. Had she activated the third law when answering the question about the hill, we might have expected confusion over inability to see a force paired with friction on the car. Instead, what we see is a very much older, "force follows velocity" style of reasoning.

Student 3, whom we will call C, did, earlier in the interview, spontaneously articulate a third law relationship. That was passed over at the time, but the interviewer returns to it at this point. The teacher's name has been suitably anonymized.

I: Earlier we were talking about laws, Newton's laws, and stuff, and you said "I don't know the numbers." Uh, and you said something about objects pushing on each other.

C: Yeah, equal but opposite forces.

I: Can you explain what that is, or just give me an example or something?



C: Ms. Frizzle did a really good one, it was when, like a huge truck and a little car and they hit each other they would each experience a force and it would be equal but it would be opposite, so as soon as they hit the force acting on it would be going backwards but it's an equal but opposite force.

I: So the car experiences a force that's the same size as the truck. . .

C: Yeah.

I: Does the truck experience a force from the car?

C: Yes, it does

I: And the forces they experience are equal?

C: Yeah, but they'd both be going opposite directions. If they hit head on, they'd experience equal but opposite forces.

I: Are there any instances with this launcher, this is the idea of forces in pairs, it's also Newton's third law that we're talking about. So are there any instances, you know, where we're seeing these equal and opposite forces being demonstrated to us?

C: Uh, [long pause] well, the only thing I can think of that would really affect it is if it hit something, like another car on the ramp or something. Um, because I'm not so sure about the force of friction, um, if that would be an equal, I don't think that's an equal and opposite force.

I: What about the launcher and the car?

C: [quickly] Yeah, the car's pushing back on the launcher and the launcher is pushing forward on the car, and the forward force of the launcher, it's got more push and that's what makes it move.



I: So the car pushes on the launcher, and the launcher pushes on the car.

C: Yeah

I: So the launcher is feeling the force from the car, and it's the same size as

the one it exerts on the car

C: Yeah

I: What about when it's on the hill? Is there any third law interaction when

it's on the hill, rolling down the hill?

C: Um, I'm not so sure. I can't think of one.

C clearly has better access than either A or B to third law relationships since he

mentioned it spontaneously as part of his analysis of an earlier event. However, he is also still

obviously struggling with applicability conditions. Collisions appear to be his primary index

to third law reasoning memories and, with prompting, he can map the collision to the action

of the spring launcher. There is a hint that he may view "equal and opposite" as a property

of an individual force rather than force pairs in an interaction, but in any case he is unable to

make the transfer to a gravitational interaction that is more dissimilar to collisions than is the

action of the launcher. Furthermore, in analyzing the launcher, in the space of just a few

seconds he passes from stating that the launcher has "more force" to arguing that forces are

equal, with no apparent trouble. That appears to be a very rapid framing change from trying

to understand how the car starts moving (there must be more force in the direction of

motion than opposite) to adopting a paired forces frame that requires a different relationship

to be articulated. The more concrete situation appears to elicit a different understanding than

the more abstract discussion that immediately follows.



Student 4, whom we will call D, also spontaneously mentioned third law relationships earlier in the interview, and the interviewer returns to that statement now.

I: We were talking a little bit about Newton's third law and forces in pairs and you talked at the beginning about normal force and gravity and you talked about in the launcher there being equal but opposite pairs. While it's on the hill, are there some equal but opposite forces?

D: Of course, gravity, normal force .. well, actually . . . wow. Well gravity's what's pulling the car down the hill . . . so . . . [stops]

I: Is it the same as forces in pairs in the launcher, or do you think it's different?

D: Different.

I: Why? Or how is it different? Explain that to me.

D: Well, in the ramp, the gravity is what's pulling the car down the ramp but in the block, the block is what's pushing the car forward.

I: But the car was pushing on the block, you said. So the ramp's pulling on the car, but do you think the car is pulling on the ramp?

D: No. The car's pushing down on the ramp, but it's not … pulling … on the ramp.

D was clearly quite capable, having spontaneously retrieved the third law and applied it correctly to the launcher at an earlier point. However, she still has some difficulty detaching this from forces that sometimes coincidentally happen to be equal and opposite,



like normal force and gravity, and appears to be in some doubt about those as a third law force pair when her attention is directed to the hill. Perhaps she suddenly realizes that in this instance the normal force is neither equal nor opposite to gravity, but she does not articulate an explanation. There's no real applicability condition issue here, but instead there is a force identification issue, and consequent problem identifying the third law partner of the gravitational force on the car. It is still an applicability problem but of a different sort than the other three students. She has ready access to the Third Law and wants to frame the hill problem that way, but cannot see how to make it work.

There are idiosyncratic issues for each of these students, but we can draw these lessons:

1. Each of them has a clear conceptual understanding of Newton's Third Law, though they could not always remember it by name, and at least in the abstract they express an understanding that the force pairs arise out of a common interaction.

2. Two of them had to be reminded of the third law and two of them used it spontaneously (recall this was at a time when there was 100% incorrect performance on the quizzes).

3. Three of them are clearly wavering on the universality of the Third Law. Sometimes they declare it to be universal, but when placed in a situation involving friction and/or normal force, none of them are bothered by the apparent lack of a force pair. The fourth believes in universality, but then is confused by a situation in which the other member of the force pair is not evident.



4. Sometimes they recognize force pairs arising from interacting objects and other times they reason backwards from accidental equality to an implication of a force pair relationship.

5. Third law knowledge has not been lost. It is clearly still present. It is also clearly not currently being interfered with, since there is no evident more recently acquired conceptual machinery being articulated at any point. The only instance of an alternative concept being substituted is when B uses the "force follows velocity" phenomenological primitive to work through the hill problem, but that is a very much older, pre-instruction, style of reasoning. The whole notion of interference depends on one conceptual reasoning structure crowding out another, and there is no evidence of that happening here.

Overall, they are struggling with applicability conditions, but in a couple of different ways. Some of them are failing to activate third law knowledge in situations where there are invisible forces acting. Others, having activated the third law knowledge, are struggling with mapping it onto the forces that they can perceive. Their responses are consistent with the pattern of incorrect responses on the RB quiz, which in turn was consistent with a belief that only active objects produce forces, so if there is only one active object there will be inevitable difficulty identifying force pairs. This is, in general, consistent with the explanation I have articulated. There don't seem to be any particular technical issues with the principal itself in most cases as they are quite willing to articulate it in an abstract sense as well as to assert universality. Rather, there is an application failure in the sense that they either don't activate the third law concept at all or they have an incomplete set of applicability conditions



for using it in a new situation. Recall, also, that at the same time their performance on the RB quiz is degrading *on questions they answered successfully before*.

Viewing this as framing, as the resources model would advocate, then it is framing on a micro level. A clear majority of these students have framed parts of this situation as a problem similar to what they have done in class and accessed appropriate resources. But they are having difficulty, for example, framing the friction situation as being in any way like the launcher situation. While this may reflect a conceptual resource involving friction that is in an intermediate state of development, one student goes so far as to reject the very universality of the force-pair picture that he vigorously asserted at a previous point. It is difficult to understand this in any terms other than constructing a third law frame that includes the launcher but excludes the friction.

In addition, it should be pointed out that there is a small amount of independent evidence supporting this viewpoint in which physical understanding is present but is inactive and replaced by older phenomenological reasoning patterns. McCaskey and Elby (2005) and Gray, et. al. (2008) performed similar experiments, the former with the FCI and the latter with the CLASS attitudinal survey. In each case, they asked participants to provide two responses – their own intuitive answer, and what they believed a physicist would say. In both experiments, they discovered a significant number of students who split their answers. These students seem to possess a conceptual understanding of Newtonian reasoning and general processes of physics, but do not, in some sense, truly believe them. That sort of response could also be interpreted as students reasoning through the applicability of concepts rather than the concepts themselves



## Discussion

### Implications of U-shaped Development for Education Research Practices

Perusing the lengthy (and, it should be noted, still incomplete) list that opens section 6 above of disparate and seemingly unrelated skills in which U-shaped development has been documented, one is led inevitably to the conclusion that it is a general feature of cognition. The exact details of the cognitive processes leading to U-shaped development are not at all important for what is to follow. It suffices that I have documented the existence of the phenomenon in the development of physical thinking. Regardless of the underlying cognitive mechanism, the mere existence of U-shaped development has profound implications both for curriculum design as well as for research practice.

As practitioners of education research, we tend to operate on the basis of two implicit underlying assumptions. One is that once students have learned something, they know it. This is probably true, but naïve, since it tends to lump complex collections of separately evolving processes into a single monolithic block.

The other, and much more insidious, assumption is that performance closely parallels understanding. It does not. We assume that once a performance change has happened, it doesn't unhappen, but that is, in fact, not the case. This is because we assume that the only cause of unhappening is loss of knowledge but, as we have seen, that is not necessarily true. The existence of U-shaped development is conclusive disproof of this assumption.

The whole enterprise of pre/post testing, by far the most widely used tool in the development and validation of physics pedagogy, is dependent on the belief that performance can be used as a reliable proxy for understanding. But we have seen that



performance and understanding are not tightly coupled, that understanding can improve while performance degrades, that this is a fundamental and general feature of cognition determined by the underlying developmental processes, and that, therefore, using performance as anything more than a rough proxy for understanding is probably not valid.

This is especially true of short time scale pre/post testing, as has often been used in the development of small scale interventions such as tutorials (see, for a small selection of examples, McDermott et. al. 1994, Wosilait et. al. 1998, Ambrose et. al. 1999, Redish, Saul and Steinberg 1997, Wittman, Steinberg and Redish 1999, Bao and Redish 2002, Heron et. al. 2003, Ortiz, Heron and Shaffer 2005). It is often difficult to determine the time scale over which these assessments are conducted based on the published data, but it often seems to be only a matter of days between pretest and post test. This means that the post tests are unlikely to have measured anything beyond an initial peak of development, and what happens after that point is generally unknown. They clearly increase conceptual understanding, but as we have seen, that is just one component of a developmental process involving multiple resources, each with its own developmental timeline. What is important, then, is the ongoing scientific enterprise within which the tutorial is embedded which will provide its context and imbue it with much of its meaning. But often the tutorial actually *is* the context and it is not clear what long term outcomes would follow.

The perceptive reader will at this point use my objections to pre/post testing to criticize the evidence on the effectiveness of LBD presented at the start of this paper which, of course, relied on pre/post testing of LBD and comparison classes. This is a point I alluded to earlier in a footnote, and I return to it at this time. Indeed, if that were the solitary line of evidence, this objection would be quite valid. However, the LBD observations



correlated multiple measures carried out over the entire time span of the unit. Ethnographic, interview, and quiz data allow us to make a reasonable interpretation of the post test data as having occurred at the end of the developmental process, and that is in turn what provides meaning to the computed gains.

**Implications of U-shaped Development for Pedagogical Design**

U-shaped development, and the process-based cognitive explanations that underlie it, provide the theoretical basis for imposing an "iterative (LBD)" or "spiraling back (Arons 1990)" structure on instruction. The reason is simple and obvious. If learning is characterized by increments followed by decrements in performance as the mental representation of that knowledge is reorganized, and if shifting context is critical to the reorganization process, then curtailing that process after observing the first peak in performance runs the risk of losing access to that knowledge altogether. This directly implies that an iterative structure should be central, not peripheral. We can go further than this. An interference explanation would impose no particular requirements on pedagogical structure other than that one should not leave material before it has become fully consolidated in memory. Spiraling is optional. It presents a tale of competition for limited cognitive resources, the solution to which is to avoid competition until the resource demands have settled to a minimal level. A multiple resources account *demands* spiraling as the only available means of broadening the range of contexts within which the appropriate concepts will be triggered into activity, thereby altering their activation priority relative to previously existing phenomenological reasoning.



This is a principle generally honored more in theory than in practice. Almost all college or high school textbooks (and therefore almost all courses) present a linear array of topics. Opportunities for spiraling back are occasional (for example, mentioning Newton's Laws again when we get to electricity), and of limited scope and duration. Spiraling is not central to the operation of the pedagogy. This is true even of research influenced texts, which often reduce or reorder the elements in the array but often do not change its linearity. If development is curvilinear, then pedagogy should be as well.

One partial exception to this rule is the Paradigms and Capstones approach pioneered by Oregon State University (Manogue et. al. 2001). This imposes an iterative structure on the entire undergraduate physics curriculum, but does not consistently do so on the component modules. We at Spelman College are currently working on implementing an LBD-like pedagogy at the college level, and in the process have expanded the design side of the LBD learning cycle to include design of models to explain physical phenomena. That work is, however, in its early stages.

Every pedagogical design, of course, embodies a conversation or negotiation between the instructor and the students, rather than simply something that is done to the students. This is one of the central points of constructivism. In that case, then, it is equally as important to think about the students' role in the conversation as the instructor's. Both of them can initiate a context shift, which I have argued is likely to be the main transformative factor in U-shaped development, and then think about how a new idea might play out in a different situation or might be connected to prior experience. This is not ordinarily something that students do habitually, though it has been observed to be a practice characteristic of working scientists (see, e.g., Dunbar, 2000). Building in opportunities,



affordances and scaffolding to create opportunities to move away from the details of the current investigation for a time and think about what other situations might embody the same concepts would seem advisable. In fact, many of the reflection tools provided to LBD students have this purpose. But many tools out of the physics education research tradition do not. That does not, of course, mean that these tools could not be embedded in larger pedagogical structures, which in fact is what we and many others do with them.

The existence of U-shaped development also has important implications for student evaluation. It implies that single point assessments are unfair and inaccurate. The data above shows not only that performance development is U-shaped but also that it is at best only partially synchronized between students. If, say, a midterm exam is scheduled at a time when one student is at the first peak of performance and another is in the valley between peaks, the more knowledgeable student can actually receive the lower grade. This in turn implies that the usual distinction between formative assessments, embedded in a unit and intended to make the state of a student's understanding visible to the student, and summative assessments, stuck at the end of the unit and assessing how well the student's knowledge state matches a goal state, is artificial and obscures important information. The embedded formative assessments, done properly, could show a developmental path and therefore provide critical contextual information for interpreting the summative assessment results. Without that contextual information, the summative measurement is a measurement of a specific performance state but you don't know which one. But considered as the end of a continuum of performance states, the overall sequence provides some opportunity to determine the stability of the underlying knowledge state. How one implements this



principle into a system of grades is not clear at present but it is an idea to which I have been giving much thought lately.

## Conclusion

What I have endeavored to show in this paper is the earliest documented evidence of U-shaped development in physics learning, together with what seems to be a plausible explanation that is consistent with all the data presented from multiple sources. That explanation is that each element of Newtonian reasoning is comprised of multiple subresources that develop at different rates. Specifically, those framing resources that lead to activation of Newtonian reasoning in memory develop more slowly than the conceptual resources of Newtonian reasoning itself. It appears that context shifting and reflection are the critical features leading to development of framing resources. Case-based reasoning provides a mechanism accounting for this differential development rate. The successive revision and abstraction of memory indices to make them fit context shifts makes them, over time, a more likely fit to novel contexts and therefore more likely to be activated. But early on, when those memories are tied to indices tightly bound to the original learning context, activation is less likely during a context shift and so performance declines. Furthermore, if as I have suggested, the atomic elements of Newtonian concepts from a developmental point of view differ from the physical point of view and, early on, are not well integrated by a web of indices, then we see partial activation during that decline.

While there may still be some room for debate over the exactly correct explanation of these and other related observations, the multiple resources explanation presented here has the following advantages:



- It accounts for how a pure context shift, with no new concepts introduced, can still cause a performance decline.

- It is a prediction of the case-based reasoning model and is at least a natural fit with the learning resources model.

- It accounts for the patterns of incorrect responses on the quiz questions, and in particular explains why the performance decline should be to a prior *intermediate* state of understanding rather than a total inability to access the concept. There generally appears to be a partial loss rather than a complete loss on the quizzes.

- The same explanation also accounts for those situations in the interview data where knowledge *is* intermittently accessible and inaccessible, depending on context.

- It accounts for why the performance decline on Second Law does not occur when Third Law was introduced but instead at a later point.

- It may explain anomalous validity results for the FCI as being due to the fact that lower level resources, from a developmental point of view, do not divide the conceptual space up in the same way as does the physicist point of view.

- It accounts for both these data and the Sayre data whereas the interference hypothesis does not.

- It easily accommodates observations of U-shaped development that play out over very long times, sometimes years, for which interference is a more problematic fit.

- It makes a testable prediction: epistemological resources should develop, in general, at a slower rate than conceptual resources.



There are several implications of the U-shaped development phenomenon, some obvious and some not. For one thing, stand alone pre/post testing, which has frequently been used for validation of physics pedagogical interventions, must be called into question since one does not know when in the developmental process the post test occurs. Gains measured by the method could be accurate, or artificially good, or artificially bad, depending on when in the developmental process the post test happened. Use of pre/post testing should likely be restricted to formative development and supported by independent lines of evidence for summative assessment.

A second implication is that for most complex conceptual issues, an iterative learning design is probably essential, not optional. Cycling back through the same set of concepts with a shifting context is the only obvious pedagogical design that matches the developmental pathway and easily affords the opportunity to detach concepts from specific contexts and elevate them to a general, abstract level. This does not, of course, mean that iteration by itself will cause this to occur. Appropriate supports must be crafted to exploit that affordance.

A less obvious point is that U-shaped development may play out over unexpectedly long periods of time. This may require a rethinking of how long a learning episode and its associated assessment should be. We should perhaps be thinking of ways to implement extended instruction/assessment procedures that may cover an entire undergraduate career, or a substantial chunk thereof.

Finally, there is the quite interesting hint that the underlying pieces of composite concepts, from a developmental point of view, do not align neatly with a physicist's viewpoint on what the component pieces should be (see figure 8 below). As mentioned



previously, I suspect this may be why instruments such as the Force Concept Inventory have validity as gestalt instruments when the validity of their individual items is problematic. But more than this, it suggests that pedagogies and developmental theories that begin with the physicist's conceptual armamentarium as atomic elements of instruction are at least partly missing the point. Certainly this is where we would want students to end up, but if they are initially operating with a much different set of tools, not exactly coinciding with any of the physicist's tools but overlapping with several of them at once, then pedagogies and theories that are not cognizant of this fact are necessarily missing a significant part of the phenomenon.

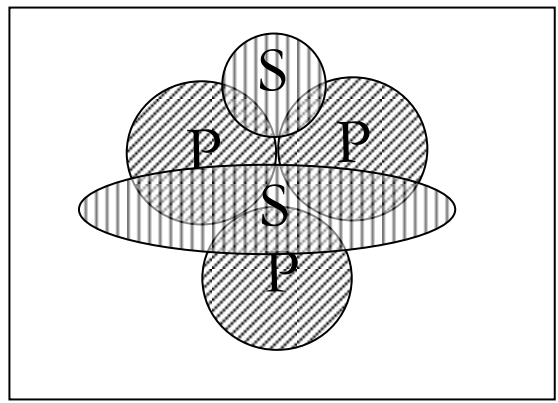

Figure 8: *Cartoon Venn Diagram of Student vs. Physicist Conceptual Resources*

A colleague of mine recently raised this question: Given the large and ever increasing amount of activity in physics education research, why are we not seeing any more large gains in understanding among our students?



This is a perceptive question, even more so when one considers that almost every pedagogical intervention ever reported has been shown to be successful by whatever measures were being used. If everything works, why are we still at the same place?

One obvious reply is that much of the field has moved from the nuts and bolts level of empirical questions about what works to more general issues of cognition without obvious immediate applications to classroom contexts. But also when physics instruction made the transition from lecture to guided interactivity, there was a discrete jump in student performance. Changes since that time that have been incremental. There are likely a variety of reasons for this. One is that changing from lecture to guided interactivity was a big sea change in the way students interact with the content, but subsequent innovations have been mostly in the line of improving that paradigm, which has become pretty much the new standard of physics pedagogy, rather than exploring its limitations and alternative paradigms.

However, the above arguments suggest a second possible factor. Since the community has historically relied on standalone pre/post testing to evaluate interventions, and since the post tests could be occurring early in the developmental process with an inevitable decline soon to follow, it is at least possible that we are abandoning instruction before the developmental process runs its course, and our students are being left in a state in which, at best, they possess fragmented, poorly indexed and often inaccessible knowledge or, at worst, they have lost access to it altogether and frequently have to start over.

Those of us whose lot in life is to teach intermediate mechanics will heave a weary sigh of recognition at this point.



## Acknowledgements

The author would like to acknowledge Noah Podolefsky, Elizabeth Charles, Eugenia Etkina, Jacquelyn Gray and Janet Kolodner for serious and productive discussions during the writing of this paper.

## Appendix: Sample Questions From the Quizzes
### Acceleration

***A-L1:*** Which of the following is **not** an **acceleration**?

a)   A change in the motion of an object.

b)   An increase in the velocity a moving object.

c)   A decrease in the velocity a moving object.

d)   An object moving rapidly.

***A-M1:*** Below are pictures of a person pulling three different wooden carts across a flat, level tabletop to the right. Each test takes 10 seconds.

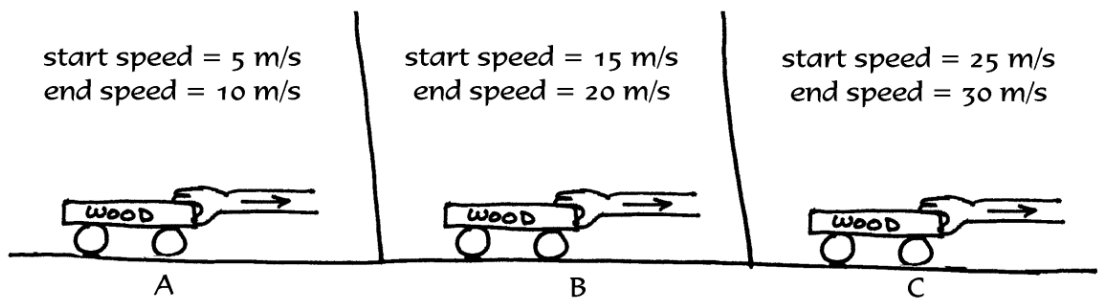

In which case is the acceleration **largest**?

a)   A has the largest acceleration.

b)   B has the largest acceleration.

c)   C has the largest acceleration.



**A-H1:** Which situation is **not** an example of **acceleration?**

a)   A ball rolling up a hill and slowing down.

b)   A ball rolling around the bottom of a bucket, at the edge, at a constant speed.

c)   A ball rolling at constant speed in a straight path across a tabletop.

d)   A ball falling from a table towards the floor.

**First Law**

**N1-L2:** Which situation shows forces that are balanced (that is, net force equal to zero, Net

Force=0)?

a)   A ball falling from a table towards the floor.

b)   A ball resting on a table.

c)   A ball rolling along a tabletop and slowing down as it rolls.

d)   None of the above.

**N1-M2:** An object is moving to the right at a constant velocity. Which of these is the best

description of the net force on the object?

a)   The net force is to the right and constant.

b)   The net force is to the right and decreasing.

c)   The net force is zero.

d)   The net force is to the left and decreasing.

e)   The net force is to the left and constant.



***N1-H2:*** When an object changes its **direction** of motion, which of the following is true?

1)  There was a change in the net force acting on the object.

2)  The object experienced an acceleration.

> a)    1 and 2 are true.
>
> b)    Only 1 is true.
>
> c)    Only 2 is true.
>
> d)    Neither is true.

**Second Law**

***N2-L2:*** A cart is pushed across the gymnasium floor, and the cart's acceleration is measured. Then, the cart is filled with a bag of sand. The same *push* (force) is given to the cart.

a)  The acceleration will be smaller for the second run.

b)  The acceleration will be larger for the second run.

c)  The acceleration is the same for both runs.

.

***N2-M3:*** [A shopping cart from a previous problem that has been given an initial push and then released] is slowing down at a steady rate. Which of the following is the best explanation for the slowing down?

> a)  The cart is slowing down because there are no forces acting on it at all, so the speed decreases.
>
> b)  The amount of "push force" from the person decreases as it coasts along, so the speed will also decrease.



c)  The friction force is decreasing at a steady rate because the speed is decreasing.

Friction slows the cart a lot at first and less later on as it coasts to a stop.

d)  The friction force is increasing to overcome the speed.

e)  The friction force is constant and creates a constant acceleration, slowing the cart at

a steady rate.

**N2-H1:** A rocket is firing its engine backward to slow itself down. As it is using up fuel, its mass of the

rocket gets smaller since the fuel is *in* the rocket. Assume that the force from the engine slowing down the

rocket remains at the same level throughout the time the rocket slows down.  When does the rocket have

the largest acceleration during this slow down?

a)  When it starts slowing down.

b)  Just before it stops.

c)  In the middle of the slowing down process.

**Third Law**

**N3-L4:** When a rocket travels through the atmosphere and in space, the push on the rocket comes from

a) the exhaust of the rocket pushing off the surface of the earth.

b) the exhaust of the rocket pushing off the air outside the back end of the rocket.

c) the exhaust of the rocket pushing on the inside of the rocket toward the direction the rocket travels.

d) none of the above.



***N3-M4:*** Question involving a moving truck colliding with a stationary truck from the FMCE. Not revealed to preserve FMCE security. The FMCE may be obtained from its authors to see this question.

***N3-H2:*** Question involving a moving car colliding with a moving truck from the FMCE. Not revealed to preserve FMCE security. The FMCE may be obtained from its authors to see this question.

***NF-L2:*** In question 2 [a friction question involving a block pulled across the table, then pulled across on wheels, then pulled across on wheels with an oiled axle], which direction does the net force act in situations A, B, and C?

a)  To the right.

b)  Down, into the tabletop.

c)  To the left.

d)  The net force in each situation is in a different and unique direction.

***NF-M2:*** A person pushing a cart, and then, shoves the cart away.  The cart continues to coast, but there is no longer any contact between it and the person.

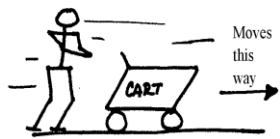

The arrows below **do not** describe the sizes of the forces involved, just their direction. **Pick the letter** of the drawing (A, B, C or D) that best describes the forces on the cart while it coasts across the floor.



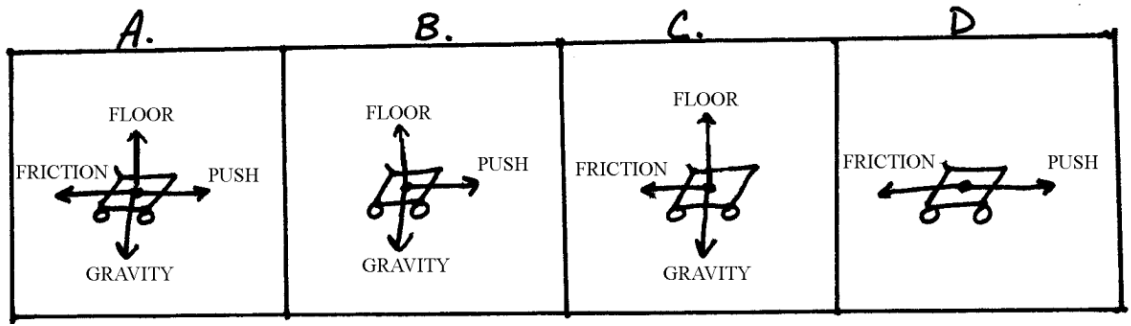

**NF-H1:** A cannonball is launched directly **straight** up into the air.

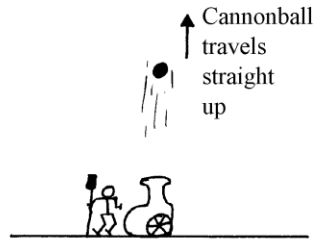

As the cannonball travels up, slowing down, the net force on the cannon ball is

a)   straight up.

b)   straight down.

c)   to the left.

to the right



References


Ambrose, B. S., Heron, P. R. L., Vokos, S. and McDermott, L. C. (1999) Student
        understanding of light as an electromagnetic wave: Relating the formalism to physical
        phenomena, *American Journal of Physics,* **67**, 10, p891

Arons, A., (1990) *A Guide to Introductory Physics Teaching,* New York, NY: Wiley and Sons

Bamberger, J. (1982) Revisiting children's drawings of simple rhythms, in Strauss op. cit.

Bao, L. and Redish, E. F. (2002) Understanding probabilistic interpretations of physical
        systems: A prerequisite to learning quantum physics, *American Journal of Physics,* **70**, 3,
        p210

Barrows, H. S., *How to Design a Problem-based Curriculum for the Pre-clinical Years,* New York,
        NY: Springer

Bouton, M. E.(1993) Context, time and memory retrieval in the interference paradigms of
        Pavlovian learning, *Psychological Bulletin*, **114**, p80

Bowerman, M. (1982) Starting to talk worse, in Strauss op. cit.

Brainerd, C. J. (2004) Dropping the other U: An alternative approach to U-shaped
        developmental functions, *Journal of Cognition and Development*, **5**, 1, p81

Brown, A. L., Ash, D., Rutherford, M., Nakagawa, K., Gordon, A. and Campione, J. (1993)
        Distributed expertise in the classroom, in G. Solomon(ed.), *Distributed Cognitions:*
        *Psychological and Educational Considerations* New York, NY: Cambridge University Press

Camp, D. (1963) My dad, private communication

Carey, S. (1982) Face perception: Anomalies of development, in Strauss op. cit.

Carroll, L. (1898) *The Hunting of the Snark: An Agony in 8 Fits*, New York, NY: Macmillan.





Clark, D. and Linn, M. C. (2003) Designing for knowledge integration: The impact of instructional time, *Journal of the Learning Sciences*, **12**, 4, p451

Clark, J. W., Sayre, E. C. and Franklin, S. V. (2010) Fluctuations in student understanding of Newton's Third Law, retrieved from http://arxiv.org, arXiv:1009.0260v1

Dawson-Tunik, T. L., Commons, M. and Fischer, K. W. (2005) The Shape of Development, *European Journal of Developmental Psychology*, **2**, 2, p163

diSessa, A. (1993) Toward an epistemology of physics, *Cognition And Instruction*, **10**, 2 & 3, p105

diSessa, A. (1996) What do "just plain folk" know about physics? In D. R. Olson and N. Torrance (eds.), *Handbook of Education and Human Development*, Oxford, UK: Blackwell

Dunbar, K. (2000) How scientists think in the real world: Implications for science education, *Journal of Applied Developmental Psychology*, 21, 1, p49-58.

Elman, J. F., Bates, E. A., Johnson, M. H., Karmiloff-Smith, A., Parisi, D. and Plunkett, K. (2001) *Rethinking Innateness: A Connectionist Perspective on Development*, London, UK: Routledge

Emmerich, W. (1982) Nonmonotonic developmental trends in social cognition, in Strauss op. cit.

Feldman, D. H. (1994) *Beyond Universals in Cognitive Development*, Norwood, NJ: Ablex

Feldman, D. H. and Fowler, R. C. (1997) Piaget, Vygotsky and the nature of developmental change, *New Ideas in Psychology*, **15**, p195

Friend, M. (2001) The transition from affective to linguistic meaning, *First Language*, **3**, p26

Gardner, H. and Winner, E. (1982) First intimations of artistry, in Strauss, op. cit.





Gershkoff-Stowe, L. and Thelen, E. (2004) U-Shaped changes in behavior: A dynamic systems perspective, *Journal of Cognition and Development*, **5**, 1, p11

Gray, K. E., Adams, W. K., Wieman, C. E. and Perkins, K. K. (2008) Students know what physicists believe, but they don't agree: A study using the CLASS survey, *Phys. Rev. ST – PER*, **4**, 020106

Hammer, D. (2000) Student resources for learning introductory physics, *American Journal of Physics – Physics Education Research Supplement*, 68(S1), pS52-S59.

Hammer, D., Elby, A., Scherr, R., and Redish, E. F. (2005) Resources, framing and transfer, in Mestre, J. (ed.) *Transfer of Learning From A Modern Multidisciplinary Perspective*, Greenwich, CT: Information Age Publishing, p89

Hebb, D. O. (1949), *The Organization of Behavior*, New York, NY: Wiley and Sons

Heron, P. R. L., Loverude, M. E., Shaffer, P. S. and McDermott, L. C. (2003) Helping students develop an understanding of Archimedes' principle II Development of research based instructional materials, *American Journal of Physics*, **71**, 11, p1188

Hestenes, D., Wells, M. and Swackhammer, G. (1992) The Force Concept Inventory, *Physics Teacher*, **30**, p141

Holbrooke, J., Gray, J., Fasse, B., Camp, P., and Kolodner, J. (2000) Assessment and evaluation of the Learning by Design physical science units, 1999-2000. Retrieved June 2010 from http://www.cc.gatech.edu/projects/lbd/htmlpubs/progress.html

Karmiloff-Smith, A. (1979) Micro- and macro-developmental changes in language acquisition and other representational systems, *Cognitive Science*, **3**, p91

Kelso, J. A. S. (1995) *Dynamic Patterns: The Self-Organization of Brain and Behavior*, Cambridge, MA: MIT Press




Klahr, D. (1982) Nonmonotone assessment of monotone development, in Straus op. cit.

Kolodner, J. (1993) *Case-based Reasoning*, San Mateo, CA: Kaufmann

Kolodner, J., Camp, P., Crismond, D., Fasse, B., Gray, J., Holbrooke, J., Puntambekar, S., and Ryan, M. (2003) Problem-based learning meets Case-based reasoning in the middle-school science classroom: Putting Learning by Design into practice, *Journal of the Learning Sciences*, **12**, 4, p495

Kolodner, J., Gray, J. and Fasse, B., (2003) Promoting transfer through case-based reasoning: Rituals and practices in Learning by Design classrooms, *Cognitive Science Quarterly.,* 3, p119

Koschmann, T., Myers, A., Feltovich, C. and Barrows, H. S. (1994) Using technology to assist in realizing effective learning and instruction: A principled approach to the use of computers in collaborative learning, *Journal of the Learning Sciences*, 3, p225

Manogue, C. A., Siemens, P. J., Tate, J., Brown, K., Niess, M. L. and Wolfer, A. J. (2001) Paradigms in physics: A new upper division curriculum, *American Journal of Physics*, **69**, 9, p978

Marcovitch, S. and Lewkowicz, D. J. (2004) U-shaped functions: Artifact or hallmark of development, *Journal of Cognition and Development*, **5**, 1, p113

Marcus, G. F. (2004) What's in a U? The shapes of cognitive development, *Journal of Cognition and Development*, **5**, 1, p119

McCaskey, T. L. and Elby, A. (2005) Probing students' epistemologies using split tasks, in J. Marx, P. Heron and S. Franklin (eds.), *2004 Physics Education Research Conference* Melville, NY: American Institute of Physics, p57



McDermott, L. C., Shaffer, P. S. and Somers, M. D. (1994) Research as a guide for teaching introductory mechanics: An illustration in the context of Atwood's machine, *American Journal of Physics*, **62**, 1, p46

McNeil, N. (2007) U-Shaped Development in Math: 7-Year-Olds Outperform 9-Year-Olds on Equivalence Problems, *Developmental Psychology,* **43**, 3, p687– 695

Mehler, J. (1982) Studies in the development of cognitive processes, in Strauss op. cit.

Moore, T. (2003) *Six Ideas that Shaped Physics,* New York, NY: McGraw-Hill

Morton, J. B. and Trehub, S. e. (2001) Children's understanding of emotion in speech, *Child Development*, **72**, p 834

Muir, D., Abraham, W., Forbes, W., and Harris, L. (1979) The ontogenesis of an auditory localization response from birth to four months of age, *Canadian Journal of Psychology*, **33**, p320

Namy, L., Campbell, A. L. and Tomasello, M. (2004) The changing role of iconicity in non-verbal symbol learning: A U-shaped trajectory in the acquisition of arbitrary gestures, *Journal of Cognition and Development.*, **5**, 1, p37

Neves, D. M. and Anderson, J. R. (1981) Knowledge compilation: Mechanisms for the automatization of cognitive skills, in Anderson, J. R. (ed.) *Cognitive Skills and Their Acquisition*, Hillsdale, NJ: Erlbaum

Ortiz, L. G., Heron, P. R. L. and Shaffer, P. S. (2005) Student understanding of equilibrium: Predicting and accounting for balancing, *American Journal of Physics,* **73**, 6, p545

Pinker, S. (1991) Rules of language, *Science*, **253**, p530

Pinker, S. and Prince, A. (1988) On language and connectionism: Analysis of a parallel distributed processing model of language acquisition, *Cognition*, **28**, p73




Redish,E. F., Saul, J. M. and Steinberg, R. N. (1997) On the effectiveness of active engagement microcomputer-based laboratories, *American Journal of Physics,* **65**, p45

Sayre, E. C. and Heckler, A. F. (2009) Peaks and decays of student knowledge in an introductory E & M Course, *Phys. Rev. ST-PER*, **5**, 013101

Schank, R. (1999) *Dynamic Memory Revisited*, New York, NY: Cambridge University Press

Schön, S. A. (1982) Intuitive thinking, in Strauss, op. cit.

Siegler, R. S. (1981) Developmental sequences within and between concepts, *Society for Research in Child Development Monographs*, **42**, 2 Serial No. 189

Siegler, R. S. and Richards, D. (1982) U-shaped behavioral curves, in Strauss op. cit.

Siegler, R. S., (2004) U-Shaped Interest in U-Shaped Development – and What it Means, *Journal of Cognition and Development,* **5**, 1, p1

Singh, C. (2008) Interactive learning tutorials on quantum mechanics, *American Journal of Physics,* **76**, 4&5, p400

Sokoloff, D. R. and Thornton, R. K. (1997) Using Interactive Lecture Demonstrations to Create an Active Learning Environment, *Physics Teacher*, **35**, pgs 340-346

Stavy, R., Strauss, S., Orpaz, N. and Carmi, C. (1982) U-shaped behavioral growth in ratio comparisons, in Strauss op. cit.

Strauss, S. (1982) (ed.), *U-shaped Behavioral Growth,* New York, NY: Academic

Thelen, E. and Fisher, D. M. (1982) Newborn stepping: An explanation for a "disappearing reflex," *Developmental Psychology*, **18**, p760

Thelen, E. and Smith, L. B. (1994) *A Dynamic Systems Approach to the Development of Cognition and Action*, Cambridge, MA: Bradford Books/MIT Press




Thornton, R. K.(1995) Conceptual dynamics: Changing student views of force and motion, in C. Bernardini, C. Tarsitani and M. Vicentini (eds.) *Thinking Physics for Teaching,* New York, NY: Plenum

Thornton, R. K. and Sokoloff, D. (1990) Learning Motion Concepts Using Real-Time Microcomputer-Based Laboratory Tools, *American Journal of Physics,* **58,** 9

Tomlinson, T. D., Huber, D. E., Rieth, C. A. and Davelaar, E. J. (2009) An interference account of cue-independent forgetting in the no-think paradigm, *Proceeding of the National Academy of Sciences*, 106, p15588

Wallace, J. G. (1982) An information processing viewpoint on nonmonotone assessment of monotone development, in Strauss op. cit.

Werker, J. F. and Tees, R. C. (1983) Developmental changes across childhood in the perception of non-native speech sounds, *Canadian Journal of Psychology*, **37**, p278

Werker, J. F., Hall, D. G. and Fais, L.(2004) Reconstruing U-shaped functions, *Journal of Cognition and Development.*, **5**, 1, p147

Wittman, M. C., Steinberg, R. N. and Redish, E. F. (1999) Making sense of how students make sense of mechanical waves, *Physics Teacher*, p1

Wosilait, K., Heron, P. R. L, Shaffer, P. S. and McDermott, L. C. (1998) Development and assessment of a research-based tutorial on light and shadow, *American Journal of Physics,* **66**, 10, p906